\documentclass[11pt]{article}
 \usepackage{amssymb,amsmath,amsfonts,amsthm,graphicx}
 \usepackage{lscape}
 \usepackage{hyperref}
 \usepackage{xcolor}

\def\be{\begin{equation}}
\def\ee{\end{equation}}
\def\bea{\begin{eqnarray}}
\def\eea{\end{eqnarray}}

\newcommand{\Sk}{{\rm\ \!S}}            
\newcommand{\Ck}{{\rm\ \!C}}           
\newcommand{\Tk}{{\rm\ \!T}}             
\newcommand{\Vk}{{\rm\ \!V}}             

\newcommand{\kk}{\kappa}

\newcommand{\bfI}{{\mathbf I}_{\kap}}

\def\k{{\kappa}}

 \def\aa{x}
\def\yy{y}
\def\dd{{\rm d}}
\def\area{\omega}

\def\auno{\beta}
\def\ados{\alpha}

\newcommand{\uu}{u}
\newcommand{\te}{\phi}

\def\k{{\kappa}}
 \def \kap{{\boldsymbol{\kappa}}}

\theoremstyle{plain}

\theoremstyle{definition}

\numberwithin{theorem}{section}
\numberwithin{equation}{section}

\begin{document}

\thispagestyle{empty}

\
\medskip
\medskip

 \vskip2cm

 \begin{center}

\noindent {\Large{\bf {Lie--Hamilton systems on  curved  spaces: \\[4pt] A geometrical approach}}}\\

\medskip
\medskip
\medskip

{\sc  Francisco J.~Herranz$^{1}$, Javier de Lucas$^{2}$ and 	Mariusz Tobolski$^{3}$}

 \end{center}

\medskip

{\small
\noindent
$^1$ Department of Physics, University of Burgos,   09001, Burgos, Spain

\noindent
$^2$ Department of Mathematical Methods in Physics, University of Warsaw, Pasteura 5, 02-093, Warszawa, Poland

\noindent
$^3$ Institute of Mathematics, Polish Academy of Sciences, ul. \'{S}niadeckich 8, 00-656, Warszawa, Poland\\

\medskip

\noindent
{\small 
 E-mail: {\rm   fjherranz@ubu.es, javier.de.lucas@fuw.edu.pl,  mtobolski@impan.pl
 }}

  \medskip

\begin{abstract}
A  Lie--Hamilton system  is a nonautonomous system of first-order ordinary differential equations describing the integral curves of a $t$-dependent vector field taking values in a finite-dimensional Lie algebra, a  Vessiot--Guldberg Lie algebra, of Hamiltonian vector fields relative to a Poisson structure.  Its general solution can be written as an autonomous function, the  superposition rule, of a generic finite family of particular solutions and a set of constants.  We pioneer the study of Lie--Hamilton systems on Riemannian spaces (sphere, Euclidean and hyperbolic plane), pseudo-Riemannian spaces (anti-de Sitter, de Sitter, and Minkowski spacetimes) as well as on semi-Riemannian spaces (Newtonian spacetimes). 
Their corresponding constants of motion and superposition rules are obtained explicitly in a geometric way. This work extends the (graded) contraction of Lie algebras to a contraction procedure for Lie algebras of vector fields, Hamiltonian functions, and related symplectic structures, invariants, and superposition rules.

  \end{abstract}

  \medskip
 \medskip

\medskip

\medskip




\section{Introduction}

A  {\it Lie system}  is a nonautonomous system  of first-order ordinary differential equations describing the integral curves of a $t$-dependent vector field taking values in a finite-dimensional Lie algebra of vector fields, a so-called  {\it Vessiot--Guldberg Lie algebra} \cite{CGM00,Dissertations}. The Lie--Scheffers theorem \cite{LS} establishes that a Lie system amounts to a nonautonomous system  of first-order ordinary differential equations whose general solution can be written as an autonomous  function, a {\it superposition rule}, of a generic family of particular solutions and some constants related to initial conditions~\cite{CGM00,Dissertations,LS,PW,CGM07,CGL11}. 

Some relevant examples of Lie systems are nonautonomous systems of first-order linear ordinary  differential equations \cite{Dissertations} and matrix Riccati equations \cite{PW}. Although most differential equations cannot be described through Lie systems \cite{Dissertations,In72}, Lie systems occur in relevant physical and mathematical problems, such as  Winternitz--Smorodinsky oscillators and Ermakov systems \cite{Dissertations},  which motivates their study (see~\cite{Dissertations,BBHLS,BHLS} for more applications). 

Lie systems admitting a Vessiot--Guldberg Lie algebra of Hamiltonian vector fields relative
to a Poisson structure \cite{IV,CMM97} are called {\it Lie--Hamilton (LH) systems}~\cite{CLS12Ham}. Although Lie systems and LH systems have been widely studied and applied, most of them are defined on a flat Euclidean space $ \mathbb{R}^n$ \cite{BBHLS, BHLS, BCHLS13Ham,GL17,WAH81}. In particular, their local classification on $ \mathbb{R}^2$ has  been recently established in~\cite{BBHLS},  starting from the classification of finite-dimensional Lie algebras of vector fields on $\mathbb{R}^2$ up to a local diffeomorphism developed by Gonz\'alez, Kamran, and Olver, the so-called {\it GKO classification}~\cite{GKP92}. Constants of motion and superposition rules for LH systems on $\mathbb{R}^2$ were studied in~\cite{BHLS}.

There exists an almost complete classification and derivation of superposition rules for complex Lie systems with primitive transitive Vessiot--Guldberg Lie algebras of  vector fields on homogeneous spaces due to Winternitz and collaborators \cite{SP84,SW84}. In spite of that, there are not many results for Lie systems possessing general real Vessiot--Guldberg Lie algebras on homogeneous spaces, which represents a much more complicated problem (cf. \cite{SP84,SW84}). Some results can be found on  one- and two-dimensional spheres~\cite{CGM00,Brockett, Komrakov, Doubrov}. Relevantly, the so-called {\it $t$-dependent projective Schr\"odinger equations} are Lie systems on a complex projective space admitting a real Vessiot--Guldberg Lie algebra of Lie symmetries of a Riemannian metric with positive constant curvature. This Vessiot--Guldberg Lie algebra also consists of Hamiltonian vector fields relative  to a symplectic structure coming from the quantum mechanical structure of the problem \cite{CCJ16}. Due to the lack of research on Lie systems in real manifolds, this paper aims to fill this gap in the literature  by classifying LH systems on  two-dimensional (2D) spaces with a Vessiot--Guldberg Lie algebra of Lie symmetries of a (possibly degenerate) metric of constant curvature by following a geometrical approach which also enables one to calculate their constants of motion and superposition rules explicitly.
 
 Section 2 surveys Lie systems on $
 \mathbb{R}^2$ possessing a Vessiot--Guldberg Lie algebra $V$ belonging to the class P$_1$ of the GKO classification~\cite{BBHLS,BHLS}. Such a Vessiot--Guldberg Lie algebra is isomorphic to the 2D Euclidean algebra  ${\mathfrak{iso}}(2)$. The vector fields of $V$ become Hamiltonian relative to a symplectic structure on $\mathbb{R}^2$. Moreover,  $V$ consists of Lie symmetries of the metric of the Euclidean plane $\mathbf{E}^2 := \mathbb{R}^2$. This allows us to obtain the corresponding superposition rules by using Euclidean and symplectic geometry~\cite{BHLS}.

To generalize the latter results to other (curved) spaces, we give  in section 3 a brief account on the nine 2D {\it Cayley--Klein (CK) spaces}~\cite{Yaglom, Ros, Gromova, Gromovb, CK2d},  which are collectively denoted by ${\mathbf
S}^2_{[\k_1],\k_2}$ where $\k_1$ and   $\k_2$ are two real parameters. 
 The   former    is  just  the  constant Gaussian curvature of the space, while the latter determines  the metric of the space through  ${\rm diag}(+1,\k_2)$.  Therefore, the  CK spaces cover the three classical Riemannian spaces of constant curvature  for $\k_2>0$ (sphere, Euclidean, and hyperbolic spaces), three pseudo-Riemannian or Lorentzian spaces for $\k_2<0$  (anti-de Sitter, Minkowski, and de Sitter spaces), as well as three semi-Riemannian or Newtonian spaces, so with a degenerate metric, for $\k_2=0$. The Euclidean plane $\mathbf{E}^2$ corresponds to the CK space ${\mathbf
S}^2_{[0],+}$.  

In section 4,  new Lie  systems on ${\mathbf
S}^2_{[\k_1],\k_2}$ are obtained by considering the Lie algebras of Lie symmetries of the metric on ${\mathbf
S}^2_{[\k_1],\k_2}$ in the so-called  {\it geodesic parallel coordinates}~\cite{HS02}; these are a natural generalization of the Cartesian coordinates to curved spaces. Next a symplectic form  is found to turn previous Lie symmetries into Hamiltonian vector fields,  so providing LH systems on ${\mathbf
S}^2_{[\k_1],\k_2}$. 

Previous new Lie systems admit Vessiot--Guldberg Lie algebras of conformal vector fields on two-dimensional manifolds. Although superposition rules for Lie systems on linear spaces admitting Vessiot--Guldberg Lie algebras of conformal vector fields can be found in \cite{BHLS,GL17,WAH81}, such results cannot effectively be applied to the Lie systems here proposed for a number of reasons. First, there exists no global diffeomorphism mapping Lie systems on manifolds to particular cases of the Lie systems treated in \cite{BHLS,GL17,WAH81} because, for instance, there is no diffeomorphism from a sphere to a linear space. As a consequence, there is no way to apply the superposition rules derived in \cite{BHLS,GL17,WAH81} to our Lie systems. Second, the Lie systems studied in \cite{GL17,WAH81} are related to Vessiot--Guldberg Lie algebras of larger dimension than those given in the present work. This causes the related superposition rules to depend on a larger number of particular solutions and to have different properties than ours \cite{Dissertations}. Finally, our approach is specially adapted to the geometry of the manifold where the Lie systems are defined. This involves the use of special spherical trigonometric functions and other techniques. This cannot be achieved through methods in \cite{BHLS,GL17,WAH81}, as they do not consider the geometry of ${\mathbf
S}^2_{[\k_1],\k_2}$.

Previous results are completed in section 5, where, firstly,  $t$-independent constants of motion are obtained by applying  the Poisson coalgebra approach introduced in~\cite{BCHLS13Ham}  and, secondly, superposition rules are deduced by making use of trigonometry on such (curved) spaces~\cite{HOS00}.  

Lie algebras of vector fields, Hamiltonian functions, and related structures appearing in sections 3, 4, and 5 are parametrized in terms of the parameters $\k_a$ $(a=1,2)$. 
Such expressions are illustrated for each specific space in tables~\ref{table1} and \ref{table2},   which summarize the main results of the paper. The cases with $\k_1=0$ and $\k_2>0$ recover known results on Euclidean LH systems on the plane \cite{BBHLS}. Moreover, this allows us to generalize graded contractions of abstract Lie algebras~\cite{Patera1, Patera2, graded}, which comprise the {\it In\"on\"u--Wigner Lie algebra contractions }corresponding to the limits $\k_a\to 0$, to Lie algebras of vector fields, Hamiltonian functions, etc. This highlights transitions among all of these known and new LH systems and their associated structures. In fact, contractions of Lie systems have only  been considered very recently in~\cite{Campoamor}, but  a systematic use covering contractions of vector fields, symplectic structures, constants of motion, and superposition rules  was still lacking. Finally, some open problems close the paper.


\section{A class of Lie--Hamilton systems on the Euclidean plane}

Let us consider the Euclidean plane $\mathbf{E}^2 := \mathbb{R}^2$  with global coordinates $\{x,y\}$ along with a real parameter $t$ and a nonautonomous system of   first-order differential equations 
\begin{equation}\label{system}
 \frac{{\rm d} x}{{\rm d} t  }=f(t,x,y), \qquad \frac{{\rm d} y}{{\rm d} t }=g(t,x,y),
\end{equation}
where $f,g:\mathbb{R}^3\rightarrow \mathbb{R}$ are arbitrary functions. 
 System (\ref{system})  is geometrically described by the $t$-dependent vector field 
\begin{equation}\label{Vect}
X:  (t,x,y)\in\mathbb{R}\times\mathbb{R}^2\ \mapsto\  f(t,x,y)\frac{\partial}{\partial x}+g(t,x,y)\frac{\partial}{\partial y}
\in {\rm T}\mathbb{R}^2.
\end{equation}
Conversely, the above $t$-dependent vector field induces a unique nonautonomous system of differential equations determining its integral curves given by (\ref{system}) (see \cite{Dissertations}). This justifies the use of $X$ to refer to both (\ref{system}) and (\ref{Vect}).  A Lie system on $\mathbb{R}^2$ is a system of the form
\be
X_t(x,y):= X(t,x,y)=\sum_{i=1}^l b_i(t)X_i(x,y),
\nonumber
\ee
where $b_1(t),\ldots,b_l(t)$ are some $t$-dependent  real functions  and $X_1,\ldots,X_l$ are vector fields  on $\mathbb{R}^2$ spanning an $l$-dimensional real Lie algebra $V$,  a  Vessiot--Guldberg Lie algebra of $X$. The {\it Lie--Scheffers Theorem} \cite{LS,PW,CGM07} states that a nonautonomous system of first-order ordinary differential equations is a Lie system if and only if its general solution can be described through a superposition rule~\cite{CGM00,Dissertations,LS,PW}.

  For our purposes and to illustrate the above concepts, we  consider        
   the $t$-dependent vector field  on $\mathbb{R}^2$ of the form
\be
X:= b_1(t)X_1+b_2(t)X_2+b_3(t)X_3 ,
\label{ab}
\ee
where $b_1(t),b_2(t),b_3(t)$ are arbitrary $t$-dependent functions and 
\be
  X_1:=\frac{\partial}{\partial x},\qquad X_2:= \frac{\partial}{\partial y}, \qquad X_3:=y   \frac{\partial}{\partial x} - x   \frac{\partial}{\partial y}.
\label{acc}
\ee
Hence, $X$ is related  to a  system of nonautonomous  first-order ordinary differential equations 
\be
\frac{\dd x}{\dd t}= b_1(t)+b_3(t) y,\qquad  \frac{\dd y}{\dd t}= b_2(t)-b_3(t) x .
\label{abb}
\ee
The previous system can be rewritten as a linear inhomogeneous complex differential equation 
$$
\frac{\dd z}{\dd t}=\bigl(b_1(t)+{\rm i}b_2(t) \bigr)-{\rm i}b_3(t)z
$$
 admitting a complex Vessiot--Guldberg Lie algebra $V_{C}=\langle \partial_z,z\partial_z\rangle_\mathbb{C}$ isomorphic to the complex affine Lie algebra $\mathfrak{Aff}(\mathbb{C})$. Meanwhile, complex Bernoulli equations \cite{Sc14} of order $\alpha\in \mathbb{R}\backslash\{1\}$, namely 
 $$
 \frac{\dd w}{\dd t}=a(t)w+b(t)w^\alpha
 $$
for arbitrary complex functions $a(t)$ and $b(t)$, possess a complex Vessiot--Guldberg Lie algebra $V_{CB}=\langle w\partial_w,w^\alpha\partial_w\rangle\simeq \mathfrak{Aff}(\mathbb{C})$. On the one hand, since affine Lie algebras of complex vector fields on the complex line are diffeomorphic, there exists a complex change of variables mapping one onto the other, namely $w^{1-\alpha}=z$. This maps (\ref{abb}), written as a complex inhomogeneous differential equation, onto a complex Bernoulli equation. Moreover, complex Bernoulli equations, as real Lie systems, admit a Vessiot--Guldberg Lie algebra given by the realification of $V_{CB}$, denoted by $V_{CB}^R$, which retrieves a result in \cite{BHLS}. In consequence, (\ref{abb}) must admit a Vessiot--Guldberg Lie algebra isomorphic to a Lie subalgebra of $V_{CB}^R$.

The   vector fields (\ref{acc})  span a  3D real  Vessiot--Guldberg Lie algebra  $V$  with commutation relations
\begin{equation}
[X_3,X_1] =X_2,\qquad  [X_3,X_2] =- X_1,\qquad  [X_1,X_2] =0.
\label{add}
\end{equation}
That is, $V$ can be written as a semidirect sum $V\simeq{\mathfrak{so}}(2) \ltimes \mathbb{R}^2\simeq \langle X_3\rangle\ltimes \langle X_1,X_2\rangle$, so  being  isomorphic to the 2D  Euclidean Lie algebra ${\mathfrak{iso}}(2)$. 
Consequently,  $X$  is called an  {\it ${\mathfrak{iso}}(2)$-Lie system}.

The Lie algebra $V$ belongs to the class P$_1$ of the GKO classification \cite{BBHLS,GKP92}. The system $X$ is therefore called a {\it $P_1$-Lie system}. Additionally, P$_1$ is also one of the 12 classes of finite-dimensional real Lie algebras of Hamiltonian vector fields on $ \mathbb{R}^2$ according to the classification performed in~\cite{BBHLS, BHLS}. Hence, $X$ admits a Vessiot--Guldberg Lie algebra of Hamiltonian vector fields with respect to a Poisson structure \cite{IV,CMM97}, and it becomes  an $ {\mathfrak{iso}}(2)$-LH  system. In particular,   the vector fields of $V$ are Hamiltonian relative to the (canonical) symplectic   form  
$$
 \omega=  \dd x\wedge  \dd y .
\label{ax}
$$
Their
 corresponding Hamiltonian functions, $h_i$, can be obtained by using the relation $\iota_{X_i}\omega={\rm d}h_i$ $(i=1,2,3)$; these can be chosen to be~\cite{BBHLS, BHLS}
 \be
h_1:=y,\qquad h_2:=-x,\qquad 	h_3:=\frac 12 (x^2+y^2) .
\label{axb}
\ee
Thus,  
$$
h_t:=b_1(t)h_1+b_2(t)h_2+b_3(t)h_3
\label{af}
$$
 is a  Hamiltonian function associated with the vector field $X_t$ given by (\ref{ab}) for every $t\in \mathbb{R}$. The linear space
 $  \langle h_1, h_2,h_3\rangle$ can be expanded to a finite-dimensional Lie algebra of functions, relative to the Poisson bracket $\{\cdot,\cdot\}_\omega$ related to $\omega$, by adding a new Hamiltonian function $h_0:=1$. In this way, 
\be
\{h_3,h_1\}_{\omega}=-  h_2 ,\quad\  \{h_3,h_2\}_{\omega}=  h_1 ,\quad\  \{h_1,h_2\}_{\omega}= h_0,\quad\  \{h_0,\,\cdot\,\}_{\omega}= 0 ,
\label{ag}
\ee
and   $ \langle h_1,h_2,h_3,h_0\rangle$ becomes a Lie algebra (with respect to $\{\cdot,\cdot\}_\omega$)  isomorphic to the centrally extended  Euclidean Lie algebra $\overline{\mathfrak{iso}}(2)$. This term is coined due to the fact that there exists an exact Lie algebra sequence
$$
\langle h_0\rangle \hookrightarrow \langle h_1,h_2,h_3,h_0\rangle \stackrel{\phi}{\rightarrow} V\simeq \mathfrak{iso}(2)
$$
where $\phi(h_i)=X_i$ and $\phi(h_0)=0$. Remarkably, every Lie algebra containing the Hamiltonian functions for the vector fields of $V$ will generate a Lie algebra isomorphic to $\overline{\mathfrak{iso}}(2)$ (cf. \cite[Corollary 5.4 and Proposition 5.5]{BBHLS}). We call this Lie algebra a  {\it LH algebra}, ${\cal H}_\omega$, for the LH system $X$.


\subsection{Constants of motion and superposition rules}

When a   nonautonomous system of first-order ordinary differential equations $X$ is shown to be a Lie system,  the Lie--Scheffers Theorem ensures that it possesses a superposition rule which can be deduced by standard (but generally cumbersome) methods~\cite{Dissertations,PW,CGM07}. If  $X$ is a  LH system, there also exists an alternative Poisson coalgebra approach, which enables one to obtain the corresponding constants of motion  (invariants) and superposition rules in an easier geometric manner.
This procedure  has recently been     formulated in~\cite{BCHLS13Ham} and extensively applied in~\cite{BHLS} to the 12 classes of LH systems on the Euclidean plane. In what follows, we review  the essentials of such a Poisson coalgebra procedure by applying it to the P$_1$-Lie system $X$ given by (\ref{acc}) (see~\cite{BHLS,BCHLS13Ham} for details).

Let $S\left(\overline{\mathfrak{iso}}(2)\right)$ be the {\it symmetric algebra} of  $\overline{\mathfrak{iso}}(2)$ \cite{Va84,CL99}, i.e. the algebra of polynomial functions on the elements of  $\overline{\mathfrak{iso}}(2)$. Let  $\{ v_0,v_1,v_2,v_3\}$ be a 
basis  of $\overline{\mathfrak{iso}}(2)$   fulfilling the   commutation relations ({\ref{ag}). The Lie bracket on $\overline{\mathfrak{iso}}(2)\subset S\left(\overline{\mathfrak{iso}}(2)\right)$ can be extended in a unique way to a Poisson bracket $\{\cdot,\cdot\}_S$ on $S\left(\overline{\mathfrak{iso}}(2)\right)$, which becomes a {\it Poisson algebra} \cite{IV,BCHLS13Ham}. Then, $S\left(\overline{\mathfrak{iso}}(2)\right)$
has a  second-order Casimir  invariant  \cite{Pa76}
\be
C:=v_{3}v_{0}-\tfrac{1}{2}(v_{1}^{2}+v_{2}^{2}),
\nonumber
\ee
i.e. $C$ is a quadratic function in the variables $v_0, v_1,v_2,v_3$, and $\{C,w\}_S=0$ for every $w\in S(\overline{\mathfrak{iso}}(2))$. The tensor product of Poisson algebras becomes a Poisson algebra in a canonic way \cite{IV}, and  $S\left(\overline{\mathfrak{iso}}(2)\right)$  can be  endowed with a {\it Poisson coalgebra} \cite{BCHLS13Ham} structure by means of the non-deformed {\it coproduct map}
${\Delta} :S\left(\overline{\mathfrak{iso}}(2)\right)\rightarrow
S\left(\overline{\mathfrak{iso}}(2)\right) \otimes S\left(\overline{\mathfrak{iso}}(2)\right)$ defined by requiring $\Delta$ to be a linear morphism such that $\Delta(wv)=\Delta(w)\Delta(v)$ for every $w,v\in S\left(\overline{\mathfrak{iso}}(2)\right)$ and 
\begin{equation}
{\Delta}(v_a):=v_a\otimes 1+1\otimes v_a,  \qquad    a=0,1,2,3,
\label{aj}
\end{equation}
namely $S\left(\overline{\mathfrak{iso}}(2)\right)$ is a Poisson algebra and $\Delta$ is a Poisson algebra homomorphism.  
 The Poisson algebra morphisms $D: S\left(\overline{\mathfrak{iso}}(2)\right) \rightarrow C^\infty(\mathbb R^2)$ and  $D^{(2)} :   S\left(\overline{\mathfrak{iso}}(2)\right)\otimes S\left(\overline{\mathfrak{iso}}(2)\right)\rightarrow C^\infty(\mathbb R^2)\otimes C^\infty(\mathbb R^2)$ defined by
\be
D( v_a):= h_a(x_1,y_1), \qquad  
D^{(2)}(v_a\otimes 1):=h_a(x_1,y_1),\qquad D^{(2)}(1\otimes v_a):=h_a(x_2,y_2),  \label{ak}
\ee
where $h_a$ are the Hamiltonian functions (\ref{axb}), lead to the  following $t$-independent  constants of motion  $F^{(1)}:= F$ and $F^{(2)}$  for the system  $X$  through the Casimir $C$ as follows (see \cite[Theorem 26]{BCHLS13Ham} for details)
\be
  F:= D(C),\qquad F^{(2)}:=  D^{(2)} \left( {\Delta}(C) \right),
\nonumber
\ee
namely~\cite{BHLS}
\be
 F=0,\qquad  F^{(2)}= \frac{1}{2}\left[
(x_{1}-x_{2})^{2}+(y_{1}-y_{2})^{2} \right] .
\label{am}
\ee
The previous functions become constant when evaluated on pairs $(x_{i}(t),y_{i}(t))$, with $i=1,2$, of particular solutions to $X$. Hence, they are first-integrals of the so-called {\it diagonal prolongation} of the $t$-dependent vector field $X$ to $(\mathbb{R}^2)^2$  \cite{Dissertations,CGM07}, namely the $t$-dependent vector field on $(\mathbb{R}^2)^2$ given by
$$
\widetilde X(t,x_1,y_1,x_2,y_2):=\sum_{i=1}^2\sum_{\alpha=1}^3b_\alpha(t)X_\alpha(x_i,y_i).
$$
By permuting $x_1\leftrightarrow x_3$, $y_1\leftrightarrow y_3$ and  $x_2\leftrightarrow x_3$, $y_2\leftrightarrow y_3$ in  $F^{(2)}$,  we find two functions $F^{(2)}_{13},F^{(2)}_{23}:(\mathbb{R}^2)^3\rightarrow \mathbb{R}$ of the form
\begin{equation}\label{an} 
\left\{
\begin{gathered}  F_{13}^{(2)}:= \tfrac{1}{2}\left[
(x_{3}-x_{2})^{2}+(y_{3}-y_{2})^{2} \right] ,\\ F_{23}^{(2)}:= \tfrac{1}{2}\left[
(x_{1}-x_{3})^{2}+(y_{1}-y_{3})^{2} \right].
\end{gathered}\right.
\end{equation}
Since the diagonal prolongation of $X$ to $(\mathbb{R}^2)^3$ is invariant under the permutation of variables, $F_{13}^{(2)}$ and $F_{23}^{(2)}$ are its first-integrals.

Since    $\partial(F^{(2)},F^{(2)}_{23})/\partial {(x_1,y_1)}\neq 0$, both constants of motion are functionally independent (the pair $F^{(2)},F^{(2)}_{13}$ is so as well).  This condition allows us to solve the system of equations
\begin{equation}
F^{(2)}= \tfrac 12 k^2_1\ge 0 ,\qquad 
 F_{23}^{(2)} = \tfrac 12 k^2_2\ge 0 ,\qquad
   F_{13}^{(2)}= \tfrac 12 k^2_3 >0,
\label{ao}
\end{equation}
in the variables $x_1,y_1$. In turn, we can write a function $
\Phi:(x_2,y_2;x_3,y_3;k_1,k_2)\in\mathbb{R}^2\times \mathbb{R}^2\times \overline{\mathbb{R}}_0^2\mapsto (x_1,y_1)\in \mathbb{R}^2
$, where  $\overline{\mathbb{R}}_0:=\{x\in \mathbb{R},x\geq 0\}$.
The theory of Lie systems \cite{Dissertations} ensures that  $\Phi$ enables us to write the general solution, $(x_1(t),y_1(t))$ to $X$ as a function of two particular solutions $(x_i(t),y_i(t))$ to $X$ and two constants, $k_1,k_2$, to be related to initial conditions as follows 
$$
(x_1(t),y_1(t))=\Phi(x_2(t),y_2(t),x_3(t),y_3(t),k_1,k_2).
$$

In particular, the system of equations (\ref{ao}) admits two solutions in the variables $x_1,y_1$ according to the signs `$\pm$'~\cite{BHLS}
\begin{equation}\label{ap}\left\{
\begin{gathered}
x_1^\pm(x_2,y_2,x_3,y_3,k_1,k_2)  = x_2+\frac{k_1^2+k_3^2-k_2^2}{2k_3^2}\, (x_3-x_2)\mp 2A\,\frac{(y_3-y_2)}{k^2_3},\\
y_1^\pm(x_2,y_2,x_3,y_3,k_1,k_2)   =y_2+\frac{k_1^2+k_3^2-k_2^2}{2k^2_3}\,(y_3-y_2)\pm 2A\,\frac{(x_3-x_2)}{k^2_3} ,\\
A=\frac 1{4}\sqrt{2(k_1^2k_2^2+k_1^2k_3^2+k_2^2k_3^2)-(k_1^4+k_2^4+k_3^4)} ,
\end{gathered}\right.
\end{equation}
where $k^2_3=(x_3-x_2)^2+(y_3-y_2)^2$. Then, it is guaranteed~\cite{CGM07}  that the above expressions give rise to superposition rules
$
\Phi_\pm:(x_2,y_2;x_3,y_3;k_1,k_2)\in\mathbb{R}^2\times \mathbb{R}^2\times \overline {\mathbb{R}}_0^2\mapsto (x^\pm_1,y^\pm_1)\in \mathbb{R}^2
$.

Remarkably, above results admit a geometrical interpretation. The vector fields (\ref{acc}) are  the infinitesimal generators of the isometries on the  Euclidean plane $\mathbf{E}^2$ relative to the standard metric $\dd x^2+\dd y^2$.   In particular, $X_3:= J_{12}$ is the generator of rotations around the origin on $\mathbf{E}^2$ (or the angular momentum), meanwhile $X_1:=P_{1}$ and $X_2:=P_{2}$ behave as infinitesimal generators of translations along the two basic axes $x$ and $y$, respectively.

Likewise, the invariants     (\ref{am}), (\ref{an}), and the superposition rules (\ref{ap}) can also be  geometrically described~\cite{BHLS}.  Let    $k_1$, $k_2$ and $k_3$  be  the Euclidean lengths of the segments $\overline{Q_1Q_2}$, $\overline{Q_1Q_3}$, and $\overline{Q_2Q_3}$ between the three  points $Q_1:=(x_1,y_1)$, $Q_2:=(x_2,y_2)$, and $Q_3:=(x_3,y_3)$ on  $\mathbf{E}^2$, respectively, which form a triangle $\triangle{Q_1Q_2Q_3}$.  Then, the invariants $F^{(2)}$,   $F_{23}^{(2)}$ and  $ F_{13}^{(2)}$ are just, in this order,   one half of the Euclidean distances   $\overline{Q_1Q_2}$, $\overline{Q_1Q_3}$, and $\overline{Q_2Q_3}$. Meanwhile  the area of the triangle ${\triangle}Q_1Q_2Q_3$ is just the constant $A$ in (\ref{ap}), which is, in fact,  the {\it Heron--Archimedes formula} for the Euclidean  area~\cite{HOS00}.


\section{Two-dimensional spaces of constant curvature}

This section provides the basic geometrical background to construct  a $(\kappa_1,\kappa_2)$-parametric family of LH systems on curved spaces along with their invariants and superposition rules.

Let us consider a two-parametric family of 3D real Lie algebras, denoted by $\mathfrak{so}_{\k_1,\k_2}(3)$}, which depends on two real parameters,  $\k_1$ and $\k_2$,  which  comprises the so-called  {\it CK  Lie algebras}~\cite{Yaglom, Ros,Gromova, Gromovb,CK2d,HS02, HOS00} or   {\it quasisimple orthogonal algebras}~\cite{casimir}.
  The structure constants of
$\mathfrak{so}_{\k_1,\k_2}(3)$ in the basis 
$\{P_1, P_2,J_{12}\}$ are given by 
\be
  [J_{12},P_1]=P_2,\qquad [J_{12},P_2]=-\k_2 P_1,\qquad [P_1,P_2]=\k_1 J_{12}  . \label{ca}
 \ee
The involutive
automorphisms $\Theta_0,\Theta_{01}:\mathfrak{so}_{\kappa_1,\kappa_2}(3)\rightarrow \mathfrak{so}_{\kappa_1,\kappa_2}(3)$, defined by imposing
\begin{equation}\nonumber
\begin{gathered}
 \Theta_0(J_{12})=J_{12},\quad \Theta_0(P_{1})=-P_{1},\quad \Theta_0(P_{2})=-P_{2}, \\[2pt]
\Theta_{01}(J_{12})=- J_{12}  ,\quad \Theta_{01}(P_{1})=P_1,\quad  \Theta_{01}(  P_{2})=-P_{2} , 
\end{gathered}
\end{equation}
diagonalize and commute among themselves. Hence, they induce a decomposition of $\mathfrak{so}_{\k_1,\k_2}(3)$ into common eigenspaces of $\Theta_0$ and $\Theta_{01}$ of the form $\mathfrak{so}_{\k_1,\k_2}(3)=E_{(1,0)}\oplus E_{(0,1)}\oplus E_{(1,1)}$, with $E_{(0,1)}:=\langle J_{12}\rangle$, $E_{(1,0)}:=\langle P_1\rangle$, $E_{(1,1)}:=\langle P_{2}\rangle$, and $E_{(0,0)}=\{0\}$. This gives rise to a $\mathbb{Z}_2\times \mathbb{Z}_2$-grading of the Lie algebra $\mathfrak{so}_{\k_1,\k_2}(3)$, i.e. $[E_{(\alpha_1,\alpha_2)},E_{(\beta_1,\beta_2)}]\subset E_{(\alpha_1+\beta_1,\alpha_2+\beta_2)}$ for every $\alpha_1,\alpha_2,\beta_1, \beta_2 \in \mathbb{Z}_2$. 
Hence, $\k_1$ and $\k_2$ are two
graded contraction parameters  determined by $\Theta_0$
and $\Theta_{01}$, respectively~\cite{graded}.
By rescaling the basis of $\mathfrak{so}_{\k_1,\k_2}(3)$ each parameter $\k_a$ $(a=1,2)$ can be reduced to either $+1$, 0 or $-1$.
The vanishment of  any $\k_a$ is   equivalent to applying an {\it In\"on\"u--Wigner contraction}~\cite{graded}.

The  automorphism $\Theta_0$ 
 gives rise to the  Cartan
decomposition:
\[
\begin{array}{lll}
\mathfrak{so}_{\k_1,\k_2}(3)={\mathfrak{h}_0}\oplus  \mathfrak{p}_0,&\quad  
{\mathfrak{h}_0}:=\langle J_{12} \rangle\simeq\mathfrak {so}_{\k_2}(2),&\quad 
{\mathfrak{p}_0}:=\langle P_{1},P_{2}\rangle ,
\end{array}
\]
where $\mathfrak{so}_{\kappa_2}(2)$ is the space of real $2\times 2$ matrices $A$  satisfying that  
$
A^T{\bf I}_{\kappa_2}+ 
{\bf I}_{\kappa_2}A=0,$ ${\bf I}_{\kappa_2}:={\rm diag}(1,\kappa_2).
$
The Lie algebra $\mathfrak{so}_{\k_1,\k_2}(3)$ is isomorphic to the matrix Lie algebra of $3\times 3$ real
matrices $M$ satisfying~\cite{CK2d}
\be
M^T \bfI\,+\bfI M=0,\qquad \bfI:={\rm diag}(1,\kappa_1,\kappa_1\kappa_2), \qquad \kap:=(\k_1,\k_2).
\label{ck2}
\ee
If $\bfI$ is not degenerate, then this space is indeed the so-called {\it indefinite orthogonal Lie algebra} $\mathfrak{so}(p,q)$, where $p$ and $q$ are the number of positive and negative eigenvalues
of the matrix $\bfI$.

In particular, the elements of the basis $\{P_1,P_2,J_{12}\}$ can be identified with the matrices
\be
P_1=-\k_1 e_{01}+e_{10}, \qquad
P_2=-\k_1\k_2 e_{02}+e_{20}, \qquad
J_{12}=-\k_2 e_{12}+e_{21} ,
\label{cd}
\ee
where $e_{ij}$ is the $3\times 3$ matrix with a single non-zero  entry 1 at row $i$
and column $j$ $(i,j=0,1,2)$. 

The elements of $\mathfrak{so}_{\kappa_1,\kappa_2}(3)$ generate by matrix exponentiation the referred to as {\it CK Lie group} ${\rm SO}_{\kappa_1,\kappa_2}(3)$. The matrix exponentials of $\{P_1,P_2,J_{12}\}$ lead to the following
  one-parametric subgroups of the CK Lie group ${\rm SO}_{\k_1,\k_2}(3)$:
\begin{equation}\label{ce}
\begin{gathered}
{\rm e}^{\alpha P_1}=\left(\begin{array}{ccc}
\Ck_{\k_1}(\alpha)&-\k_1\Sk_{\k_1}(\alpha)&0 \cr 
\Sk_{\k_1}(\alpha)&\Ck_{\k_1}(\alpha)&0\cr 
0&0&1
\end{array}\right) ,\qquad
{\rm e}^{\gamma J_{12}}=\left(\begin{array}{ccc}
1&0&0\cr 
0&\Ck_{\k_2}(\gamma)&-\k_2\Sk_{\k_2}(\gamma)\cr
0&\Sk_{\k_2}(\gamma)&\Ck_{\k_2}(\gamma)
\end{array}\right) , \\[4pt] 
{\rm e}^{\beta P_2}=\left(\begin{array}{ccc}
\Ck_{\k_1\k_2}(\beta)&0&-\k_1\k_2\Sk_{\k_1\k_2}(\beta)\cr 
0&1&0\cr 
\Sk_{\k_1\k_2}(\beta)&0&\Ck_{\k_1\k_2}(\beta)
\end{array}\right) ,
\end{gathered}
\end{equation}
where the so-called {\it $\k$-dependent cosine} and {\it sine} functions read~\cite{CK2d, HS02,HOS00}:
\begin{equation}
\Ck_{\k}(\uu):=\sum_{l=0}^{\infty}(-\k)^l\frac{\uu^{2l}} 
{(2l)!}=\left\{
\begin{array}{ll}
  \cos {\sqrt{\k}\, \uu} &\quad  \k>0 \\ 
\qquad 1  &\quad
  \k=0 \\
{\rm ch}\, {\sqrt{-\k}\, \uu} &\quad   \k<0 
\end{array}\right.  ,
\nonumber
\end{equation}
\begin{equation}
   \Sk{_\k}(\uu) :=\sum_{l=0}^{\infty}(-\k)^l\frac{\uu^{2l+1}}{ (2l+1)!}
= \left\{
\begin{array}{ll}
  \frac{1}{\sqrt{\k}} \sin {\sqrt{\k}\, \uu} &\quad  \k>0 \\ 
\qquad \uu  &\quad
  \k=0 \\ 
\frac{1}{\sqrt{-\k}} {\rm sh}\, {\sqrt{-\k}\, \uu} &\quad  \k<0 
\end{array}\right.  .
\label{ccj}\nonumber
\end{equation}
From them, the {\it  $\k$-tangent} and the {\it $\k$-versed sine} (or versine) take the form
\be
\Tk_{\k}(u) :=\frac{\Sk_\k(u)}{ \Ck_\k(u)} ,\qquad \Vk_{\k}(u) :=\frac 1\k \left(1-\Ck_\k(u) \right)    .
\label{cj}
\ee
These $\k$-functions cover both the usual circular $(\k>0)$ and hyperbolic $(\k<0)$ trigonometric functions. In the case $\k=0$, the previous functions reduce to the parabolic ones $\Ck_{0}(\uu)=1$, $   \Sk{_0}(\uu) =   \Tk{_0}(\uu) =u$, and $\Vk_{0}(u) =u^2/2$.

Some  relations  for the above $\k$-functions read
 \be
 \Ck^2_\k(\uu)+\k\Sk^2_\k(\uu)=1,\qquad  \Ck_\k(2\uu)= \Ck^2_\k(\uu)-\k\Sk^2_\k(\uu), \qquad \Sk_\k(2\uu)= 2 \Sk_\k(\uu) \Ck_\k(\uu) ,
\label{za}\nonumber
\ee
and their derivatives are given by
\be
 \frac{ {\rm d}}{{\rm d} \uu}\Ck_\k(\uu)=-\k\Sk_\k(\uu),\quad\        \frac{ {\rm d}}
{{\rm d} \uu}\Sk_\k(\uu)= \Ck_\k(\uu)  ,\quad\    
\frac{ {\rm d}}
{{\rm d} \uu}\Tk_\k(\uu)=  \frac{1}{\Ck^2_\k(\uu) } ,\quad\    
\frac{ {\rm d}}
{{\rm d} \uu}\Vk_\k(\uu)=  {\Sk_\k(\uu) }.
\label{zb}
\ee
Many other relations can be found in~\cite{HOS00}.

Let $H_0:= {\rm  SO}_{\k_2}(2) $  be the Lie subgroup of ${\rm SO}_{\kappa_1,\kappa_2}(3)$ obtained by matrix exponentiation of the Lie algebra  ${\mathfrak{h}_0}$.  The CK family of  2D homogeneous
spaces  is defined by  the quotient 
\be
{\mathbf
S}^2_{[\k_1],\k_2} := {\rm  SO}_{\k_1,\k_2}(3)/{\rm  SO}_{\k_2}(2).
\label{cc}
\ee
The (possibly degenerate) metric defined by   ${\bfI}$ (\ref{ck2}) on $T_e{\rm SO}_{\k_1,\k_2}(3)\simeq \mathfrak{so}_{\k_1,\k_2}(3)$ can be extended by right translation to a metric on the whole $SO_{\k_1,\k_2}(3)$ and then projected onto ${\mathbf
S}^2_{[\k_1],\k_2}$. Then, the CK family becomes a symmetric space relative to the obtained metric. The contraction parameter $\k_1$ becomes the constant (Gaussian) {\em curvature} of the space. The second parameter $\k_2$ determines the {\em signature} of the metric through ${\rm diag}(+,\k_2)$.

\subsection{Ambient, geodesic    parallel and geodesic polar coordinates}

The matrix realization (\ref{ce}) enables us to identify the elements of ${\rm SO}_{\k_1,\k_2}(3)$ with isometries of the bilinear form  $\bfI$ (\ref{ck2}). More specifically, given a $3\times 3$ matrix  $g$, it follows that 
$$
g\in {\rm SO}_{\k_1,\k_2}(3) \Rightarrow g^T \bfI\, g=\bfI.
\label{ck}
$$
This allows us to consider the Lie group action of ${\rm SO}_{\k_1,\k_2}(3)$ on $\mathbb{R}^3$ as isometries of $\bfI$.

The subgroup $  {\rm SO}_{\k_2}(2)=\langle {\rm e}^{\gamma J_{12}} \rangle$ is the isotropy subgroup of the point
$O:=(1,0,0)$, which is taken as the {\em origin} in the space
$\mathbf S^2_{[\k_1],\k_2}$. Hence, ${\rm SO}_{\k_1,\k_2}(3)$ becomes an isometry   group of the space $\mathbf S^2_{[\k_1],\k_2}$, in such a manner that  $J_{12}$ is a  rotation generator, while $P_1$ and $P_2$ move $O$ along two basic geodesics $l_1$ and $l_2$, which are orthogonal at $O$, so behaving as translation generators (see figure~\ref{figure1}).

 The orbit of $O$ is contained in the submanifold given by $\bfI$ of the form
\begin{equation}
\Sigma_\kap :=\{v:=(x_0,x_1,x_2)\in \mathbb{R}^3: \bfI(v,v)=\ x_0^2+\k_1  x_1^2+  \k_1\k_2 x_2^2    =1\} .
\label{cl}
\end{equation}
This orbit, namely the connected component of $\Sigma_\kap$ containing the point $O$, can be identified with the space  ${\mathbf
S}^2_{[\k_1],\k_2}$. The coordinates $\{x_0,x_1,x_2\}$ on $\mathbb{R}^3$,
  satisfying  the constraint (\ref{cl}) on $\Sigma_\kap$,  are called {\em ambient} or {\em Weierstrass coordinates}. In these variables, the metric on ${\mathbf
S}^2_{[\k_1],\k_2}$
comes from    the flat ambient metric in $\mathbb R^{3}$ divided by the curvature $\k_1$ and
restricted to $\Sigma_\kap$, namely
\begin{equation}
{\rm d} s_\kap^2:=\left.\frac {1}{\k_1}
\left({\rm d} x_0^2+   \k_1 {\rm d} x_1^2+   \k_1 \k_2{\rm d} x_2^2 
\right)\right|_{\Sigma_\kap}  =    \frac{\k_1\left(x_1{\rm d} x_1 +\k_2 x_2{\rm d} x_2 \right)^2}{1-  \k_1  x_1^2-   \k_1  \k_2 x_2^2 }+  {\rm d} x_1^2+   \k_2{\rm d} x_2^2 .
\label{cm}
\end{equation}
It is worth noting that if $\k_1=0$, then $\Sigma_\kap$ is given by two connected components with $x_0\in \{-1,1\}$ and ${\rm d}s_\kap^2$ is well-defined.

The ambient coordinates  can be parametrized on $\Sigma_\kap$ in terms of two intrinsic variables in different ways (see e.g.~\cite{HS02,anisotropic}). In particular, let us introduce the so-called {\em  geodesic parallel} $\{x,y\}$ and  {\em geodesic polar} $\{r,\phi\}$ coordinates of a point $Q:=(x_0,x_1,x_2)$ in $\mathbf S^2_{[\k_1],\k_2}$ which  are obtained through the following action of the  one-parametric subgroups (\ref{ce}) on $O$~\cite{HS02}:
$$
(x_0,x_1,x_2)^T = \exp(xP_1) \exp(yP_2)O^T= \exp(\phi J_{12}) \exp(r P_1)O^T ,
\label{cn}
$$
yielding
\bea
&& x_0=\Ck_{\kk_1}(x)\Ck_{\kk_1\k_2}(y)=\Ck_{\kk_1}(r),\nonumber\\[2pt]
&& x_1=\Sk_{\kk_1}(x)\Ck_{\kk_1\k_2}(y)=\Sk_{\kk_1}(r)\Ck_{\k_2}(\te) , \nonumber\\[2pt]
&& x_2=\Sk_{\kk_1\k_2}(y)=\Sk_{\kk_1}(r)\Sk_{\k_2}(\te) .
\label{co}
\eea
By introducing these relations in the metric (\ref{cm}) and applying (\ref{zb}), we recover the usual (curved) metrics  given by
\be
{\rm d} s_\kap^2=\Ck^2_{\k_1\k_2}(y){\rm d} x^2 + \k_2{\rm d} y^2   =      {\rm d} r^2+\k_2  \Sk^2_{\k_1}(r)  {\rm d} \phi^2 .
\label{cp}
\ee
As shown in figure~\ref{figure1}, the variable $r$ is the   distance   between the origin   $O$ and the point $Q$ measured along  the geodesic $l$ that joins  both points, while $\phi$ is the   angle  of $l$ relative to basic geodesic  $l_1$. 
If  $Q_1$ denotes the intersection point of $l_1$ with its orthogonal geodesic $l_2'$ through $Q$,  then    $x$  is the geodesic distance between $O$ and $Q_1$ measured  along $l_1$  and    $y$ is the  geodesic distance   between $Q_1$ and $Q$ measured  along $l_2'$.  
Note that   a second set of geodesic parallel coordinates $\{x',y'\}$, similar to $\{x,y\}$,  can also be defined by considering   
   the intersection point  $Q_2$ of $l_2$ with its orthogonal geodesic $l_1'$ through $Q$, and  that $\{x,y\}\ne \{x',y'\}$ if the curvature $\k_1\ne 0$~\cite{HS02}. On the flat Euclidean plane 
 $ {\mathbf E}^2$ with $\k_1=0$, $\{x,y\}=\{x',y'\}$ reduce to Cartesian coordinates and $\{r,\phi\}$ to the usual polar ones.


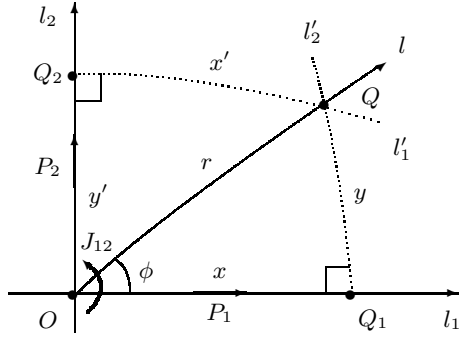
\begin{figure}[t]
\begin{center}
\begin{picture}(170,125)
\footnotesize{
\put(127,22){$\bullet$}
\put(52,33){\makebox(0,0){$\te$}}
 \qbezier(46,25)(46,35)(40,38)
 \put(80,113){\makebox(0,0){$x'$}}
\put(34,62){\makebox(0,0){$y'$}}
\put(133,62){\makebox(0,0){$y$}}
\put(80,33){\makebox(0,0){$x$}}
\put(75,74){\makebox(0,0){$r$}}
\put(142,112){\vector(4,3){1}}
\put(117,94){$\bullet$}
\put(138,15){\makebox(0,0){$Q_1$}}
\put(137,99){\makebox(0,0){$Q$}}
\put(149,80){\makebox(0,0){$l'_1$}}
\put(149,119){\makebox(0,0){$l$}}
\put(22,22){$\bullet$}
\put(15,15){\makebox(0,0){$O$}}
\put(22,105){$\bullet$}
\put(15,108){\makebox(0,0){$Q_2$}}
\put(15,130){\makebox(0,0){$l_2$}}
\put(25,10){\vector(0,1){125}}
\put(0,25){\vector(1,0){170}}
\put(168,15){\makebox(0,0){$l_1$}}
\put(35,98){\line(0,1){10}}
\put(25,98){\line(1,0){10}}
\put(120,25){\line(0,1){10}}
\put(120,35){\line(1,0){9}}
\qbezier[50](25,108)(70,110)(140,90)
\qbezier[50](130,25)(125,80)(115,114)
\put(115,124){\makebox(0,0){$l'_2$}}
\qbezier(25,24)(50,50)(140,111)
\linethickness{1pt}
\put(70,25){\vector(1,0){20}}
\put(25,65){\vector(0,1){20}}
\put(15,73){\makebox(0,0){$P_{2}$}}
\qbezier(30,18)(40,28)(30,36)
\put(29,37){\vector(-1,1){1}}
\put(75,15){$P_{1}$}
\put(27,42){$J_{12}$}
}
\end{picture}
\end{center}
\noindent
\\[-40pt]
\caption{\footnotesize Schematic representation of  the isometry infinitesimal generators $\{ J_{12},P_1,P_2\}$ and  geodesic coordinates   $\{x,y\}$, $\{x',y'\}$  and $\{r,\te\}$    of a   point  $Q=(x_0,x_1,x_2)$   on a  2D CK    space.}
\label{figure1}
\end{figure}


 Since we are interested in extending the Euclidean P$_1$-LH systems of section 2 to all the CK spaces,  we shall make use of the geodesic parallel coordinates $(x,y)$, although the relations (\ref{co})  would enable one to express our  final results in terms of the geodesic polar ones.

 Summing up, according to the values of the two $\k_a$ parameters,  the CK space $\mathbf S^2_{[\k_1],\k_2}$ comprises   {\em nine} 
 specific 2D symmetrical homogeneous spaces, which depending on  the parameter $\k_2$  are classified into three types:
 
 \begin{itemize}
 
 \item {\em Riemannian spaces} for $\k_2>0$. The standard sphere $\mathbf {S^2}$ arises when  $\k_1>0$. The case $\k_1<0$ leads to a two-sheeted hyperboloid. We call $\mathbf {H^2}$  the upper sheet of the hyperboloid, namely the part with $x_0\geq 1$: the so-called {\it Lobachevsky space}.
The contraction $\k_1= 0$   gives rise to two Euclidean planes $x_0=\pm 1$. We will call Euclidean space, $\mathbf {E^2}$, the one with $x_0=+1$.

  \item {\em Pseudo-Riemannian spaces or Lorentzian spacetimes} for $\k_2<0$.  
  For Gaussian curvature  $\k_1>0$, we obtain the {\it 2D co-hyperbolic space} or {\it $(1+1)$D anti-de Sitter spacetime} $\mathbf {AdS^{1+1}}$;  if $\k_1<0$, we find the {\it 2D doubly-hyperbolic space} or {\it $(1+1)$D  de Sitter spacetime $\mathbf {dS^{1+1}}$}; and   the flat case with $\k_1=0$ provides the $(1+1)$D  Minkowskian spacetime $\mathbf {M^{1+1}}$. In all cases for $\k_2<0$, the $J_{12}$, $P_1$, and $P_2$  correspond to the infinitesimal generators of boosts, time translations, and spatial translations, respectively. From a physic viewpoint, the $\k_a$ parameters are related to the cosmological constant $\Lambda$ and the speed of light $c$ through 
  \be
  \k_1=-\Lambda,\qquad \k_2=-1/c^2.
  \nonumber
  \ee
And the geodesic parallel coordinates $(x,y)$ are just the  time $t$ and space $y$ ones.

  \item {\em Semi-Riemannian spaces or Newtonian spacetimes} for $\k_2=0$ $(c=\infty)$. In this case, the metric  (\ref{cm}) is degenerate and the kernel of the metric gives rise to an integrable foliation of $\mathbf S^2_{[\k_1],0}$, which is invariant under the 
  action of  the CK group   ${\rm SO}_{\k_1,0}(3)$ on  $\mathbf S^2_{[\k_1],0}$.  There appears a well-defined subsidiary metric  ${\rm d} {s'}^2 :={\rm d} s_\kap^2/\k_2 $  restricted to each leaf, which in the coordinates $(x,y)$  read~\cite{HS02}
  \be
  {\rm d} s^2= {\rm d} x^2 ,\qquad   {\rm d}{s'}^2= {\rm d} y^2 \quad {\rm{on}} \quad x= \, {\rm constant} .
  \nonumber
  \ee
For $\k_1>0$ we find the {\it 2D co-Euclidean space} or {\it $(1+1)$D oscillating Newton--Hook (NH) spacetime} $\mathbf {NH_+^{1+1}}$, and for $\k_1<0$ we obtain the 
 2D {\it co-Minkowskian space} or {\it $(1+1)$D expanding NH spacetime} $\mathbf {NH_-^{1+1}}$. The flat space with $\k_1=0$ is just the Galilean one $\mathbf {G^{1+1}}$. Hence, in these three cases,  the metric 
  $ {\rm d} s_\kap^2$ provides `absolute-time' $t$, the leaves of the invariant foliation are the `absolute-space' at $t=t_0$ and   ${\rm d}{s}'_\kap{^{2}}$ is the subsidiary spatial metric  defined on each leaf.

 \end{itemize}
 
 Each specific CK space, Lie algebra of infinitesimal symmetries, and metric are displayed in table~\ref{table1} in the next section.


\section{A class of LH systems on curved spaces}

We shall hereafter make extensive use of the shorthand notation  $\kap:=(\k_1,\k_2)$. Our procedure consists in defining a Lie system $X_{\kap}$ possessing a Vessiot--Guldberg Lie algebra $V_{\kap}$ consisting of infinitesimal symmetries of the metric of the CK space ${\mathbf
S}^2_{[\k_1],\k_2}$. Next, we obtain a compatible symplectic form $\omega_\kap$ turning the elements of $V_{\kap}$ into Hamiltonian vector fields. 

The fundamental vector fields of the Lie group action of ${\rm SO}_{\kap}(3)$ on $\mathbb{R}^3$ by isometries of $\bfI$ are Lie symmetries of ${\rm d}s^2_{\kap}$. Since the action is linear, the fundamental vector fields can be obtained straightforwardly from the 3D  matrix representation (\ref{cd}).  In ambient coordinates $(x_0,x_1,x_2)$, they read~\cite{HS02},
\be
P_1:=\k_1 x_1 \frac{\partial}{\partial x_0}   - x_0 \frac{\partial}{\partial x_1} , \qquad
P_2:=\k_1\k_2 x_2 \frac{\partial}{\partial x_0}   - x_0 \frac{\partial}{\partial x_2} , \qquad
J_{12}:=\k_2 x_2 \frac{\partial}{\partial x_1}   - x_1 \frac{\partial}{\partial x_2}.
\label{da}\nonumber
\ee
Since the function ${\bfI}(v,v)=x_0^2+\k_1x_1^2+\k_1\k_2x_2^2$ is an invariant of the action of the Lie group action ${\rm SO}_\kap(3)$, the above vector fields   can be restricted to $\Sigma_\kap$. Such restrictions are Lie symmetries of the restriction of ${\rm d}s^2_\kap$. These vector fields are the  `curved'  counterpart  of the initial  Euclidean ones $X_i$  (\ref{acc}) in any coordinate system. In terms of geodesic parallel coordinates (\ref{co})  and   using   (\ref{zb}), they become   
\begin{equation}
\begin{gathered}
X_{\kap,1}:= - P_1=\frac{\partial}{\partial x}  ,\qquad X_{\kap,2}:= -P_2=\kk_1\kk_2\Sk_{\kk_1}(x)\Tk_{\kk_1\kk_2}(y)\frac{\partial}{\partial x}  +\Ck_{\kk_1}(x)\frac{\partial}{\partial y}  ,\\
 X_{\kap,3}:= J_{12}= \kk_2\Ck_{\kk_1}(x)\Tk_{\kk_1\kk_2}(y)\frac{\partial}{\partial x}  -\Sk_{\kk_1}(x)\frac{\partial}{\partial y}   .
\label{db}
\end{gathered}
\end{equation}
Then      the $t$-dependent vector field 
\be
X_\kap:= b_1(t)X_{\kap,1}+b_2(t)X_{\kap,2}+b_3(t)X_{\kap,3} ,
\label{dc}
\ee
provides the following system   of   nonautonomous   differential equations  
\bea
&&\frac{\dd x}{\dd t}= b_1(t)+ \kk_1\kk_2\, b_2(t) \Sk_{\kk_1}(x)\Tk_{\kk_1\kk_2}(y)+   \kk_2\, b_3(t) \Ck_{\kk_1}(x)\Tk_{\kk_1\kk_2}(y), \nonumber\\[2pt]
&&  \frac{\dd y}{\dd t}= b_2(t) \Ck_{\kk_1}(x) - b_3(t)   \Sk_{\kk_1}(x) .
\label{de}
\eea
Obviously, $X_\kap$ is a Lie system and   the vector fields (\ref{db})   satisfy  the    commutation relations (\ref{ca}), that is,
\begin{equation}
[X_{\kap,3,}, X_{\kap,1}] = X_{\kap,2} ,\qquad  [ X_{\kap,3} X_{\kap,2}] =- \k_2 X_{\kap,1} ,\qquad  [ X_{\kap,1} , X_{\kap,2} ] =\k_1  X_{\kap,3},
\label{df}
\end{equation}
so spanning  a Vessiot--Guldberg Lie algebra  $V_\kap$ isomorphic to the CK Lie algebra $\mathfrak{so}_{\kap}(3)$.
If we now  consider the Euclidean space  $ {\mathbf E}^2$ with  parameters $\kap=(\k_1,\k_2)=(0,+1)$, we find that vector fields (\ref{db}), differential equations (\ref{de}),  and commutation rules (\ref{df}) reduce to (\ref{acc}), (\ref{abb}) and (\ref{add}), respectively.

Furthermore,  the  restriction of the vector fields  $ X_{\kap,i}$ to $\Sigma_\kap$ can be turned into Hamiltonian vector fields with Hamiltonian functions $h_{\kap,i}$ with respect to a  symplectic form $\omega_\kap$. Recall that they are infinitesimal symmetries of ${\rm ds}^2_{\kap}$. If ${\rm ds}^2_{\kap}$ is not degenerate, then $V_\kap$ becomes a Lie algebra of Killing vector fields relative to ${\rm ds}^2_{\kap}$. Hence, they are Lie symmetries of the volume form $\omega_{\kap}$ on $\Sigma_\kap$ induced by ${\rm ds}^2_\kap$. Up to a non-zero proportional constant, 
\be 
\omega_\kap=\Ck_{\kk_1\kk_2}(y) \, {\rm d}x\wedge{\rm d}y.
\label{dd}
\ee
The case when ${\rm d}s^2_\kap$ is degenerate can be obtained by making an appropriate limit in $\kap$. Obviously, $\omega_\kap$ is the  area   element  $\dd A$ for all the CK spaces~\cite{HS02}.

Next, the relation $\iota_{X_{\kap,i}}\omega_\kap={\rm d}h_{\kap,i}$ allows us to determine some Hamiltonian functions $h_{\kap,i}$ for the vector fields $X_{\kap,i}$ with respect to the symplectic form $\omega_\kap$:
\begin{equation}
\begin{gathered}
 h_{\kap,1}=\Sk_{\kk_1\kk_2}(y),\qquad h_{\kap,2}=-\Sk_{\kk_1}(x)\Ck_{\kk_1\kk_2}(y),\\[2pt]
 h_{\kap,3}=\frac{1}{\kk_1}\bigl(1-\Ck_{\kk_1}(x)\Ck_{\kk_1\kk_2}(y) \bigr) =\Vk_{\kk_1}(x)+\k_2 \Vk_{\kk_1\kk_2}(y) -\k_1\k_2 \Vk_{\kk_1}(x)\Vk_{\kk_1\kk_2}(y) .
\end{gathered}
\label{hamfun}
\end{equation}
The above functions span, along with a function $h_{\kap,0}=1$, a Lie algebra of functions relative to the Poisson bracket $\{\cdot,\cdot\}_{\omega_\kap}$ induced by $\omega_\kap$. In fact, the base of such a Lie algebra satisfies the following commutation relations
\be
\begin{gathered}
\{h_{\kap,3}, h_{\kap,1}\}_{\omega_\kap}=-h_{\kap,2},\qquad \{h_{\kap,3}, h_{\kap,2}\}_{\omega_\kap}= \kk_2 h_{\kap,1}, \\[2pt]
\{h_{\kap,1},h_{\kap,2}\}_{\omega_\kap}=h_{\kap,0}-\kk_1 h_{\kap,3},\qquad  \{h_{\kap,0},\cdot\}_{\omega_\kap}=0.
\end{gathered}
\label{dg}
\ee
Indeed,    $h_{\kap,0}$ is a central generator in such a manner that $(\langle h_{\kap,1}, h_{\kap,2}, h_{\kap,3}, h_{\kap,0}\rangle, \{\cdot,\cdot\}_{\omega_\kap})$ span a LH algebra  ${\cal H}_{\omega_\kap}$ which is isomorphic to a central 
extension of the   CK Lie algebra $ {\mathfrak{so}}_\kap(3)$, denoted by $\overline{\mathfrak{so}}_\kap(3)$.  In this way, we obtain the $t$-dependent Hamiltonian associated with the Lie system (\ref{dc}):
\be
h_\kap(t)=b_1(t)h_{\kap,1}+b_2(t)h_{\kap, 2}+b_3(t)h_{\kap,3} .
\label{dh}\nonumber
\ee

 We remark that the addition of a central generator $h_{\kap,0}$ is necessary to ensure that  the  {Hamiltonian} functions $h_{\kap,i}$ span a Lie algebra, similarly to  the Euclidean case described in section 2. However, it is well-known that the central extension is trivial  when $\k_1\ne  0$~\cite{Azcarraga}. This, in turn, means that if we apply the change of basis
\be
 h'_{\kap,1}= h_{\kap,1},\qquad  h'_{\kap,2}= h_{\kap,2},\qquad    h'_{\kap,3}= h_{\kap,3} - h_{\kap,0}/\k_1 ,\qquad  \k_1\ne 0,
 \label{dh2} \nonumber
\ee
the trivial extension is `removed' and the commutation relations (\ref{dg}) become
\be
\{h'_{\kap,3}, h'_{\kap,1}\}_{\omega_\kap}=-h'_{\kap,2},\qquad \{h'_{\kap,3}, h'_{\kap,2}\}_{\omega_\kap}= \kk_2 h'_{\kap,1} \qquad \{h'_{\kap,1},h'_{\kap,2}\}_{\omega_\kap}=- \k_1 h'_{\kap,3}, 
\label{dgb} \nonumber
\ee
which are just  the commutation relations  (\ref{df}) of  the CK  Lie algebra $ {\mathfrak{so}}_\kap(3)$ for $\k_1\ne 0$. In this case, the LH algebra  ${\cal H}_{\omega_\kap}\simeq \overline{ {\mathfrak{so}}}_\kap(3) \simeq { {\mathfrak{so}}}_\kap(3) \oplus \mathbb R$.	
On the contrary, if $\k_1=0$ the central extension $h_{\kap,0}$  is a non-trivial one~\cite{Azcarraga} (this cannot be `removed'  through a change of basis)  and  the commutation rules  (\ref{dg}) read
\be
 \{h_{\kap,3}, h_{\kap,1}\}_{\omega_\kap}=-h_{\kap,2},\quad \{h_{\kap,3}, h_{\kap,2}\}_{\omega_\kap}= \kk_2 h_{\kap,1}, \quad 
  \{h_{\kap,1},h_{\kap,2}\}_{\omega_\kap}=h_{\kap,0}   ,\quad  \{h_{\kap,0},\cdot\}_{\omega_\kap}=0,
\label{dgc} \nonumber
\ee
which correspond to central extensions of non-simple Lie algebras: Euclidean $\overline{ {\mathfrak{iso}}}(2) \simeq{\overline{{ \mathfrak{so}}(2) \ltimes \mathbb{R}^2}}$ $(\k_2>0)$ (so recovering (\ref{ag})), Poincar\'e  $\overline{ {\mathfrak{iso}}}(1,1) \simeq{ \overline{ {\mathfrak{so}}(1,1) \ltimes \mathbb{R}^2}}$  $(\k_2<0)$, and Galilei $\overline{ {\mathfrak{iiso}}}(1) \simeq{  \overline{ {\mathbb R}\ltimes \mathbb{R}^2}}$  $(\k_2=0)$.

Notice also that the Hamiltonian function  $h_{\kap,3}$ (\ref{hamfun}) is written in two forms. The former requires to take the limit $\k_1\to 0$ for the flat cases taking power series of $\Ck_\k(u)$, but the latter (in terms of  $\k$-versed sines (\ref{cj})) directly provides  the same result by setting $\k_1=0$.

 We display in table~\ref{table1} the specific vector fields (\ref{db}), Hamiltonian functions (\ref{hamfun}), and symplectic form (\ref{dd}) for each of the nine spaces comprised within the CK family (\ref{cc}).

\begin{table}[htbp]
{\footnotesize
 \noindent
\caption{{\small LH algebras on the nine CK spaces according to the `normalized'  values of the contraction parameters  $\k_a\in\{1, 0, -1\}$. For each   space ${\mathbf
S}^2_{[\k_1],\k_2}$ (\ref{cc})  it is shown, in geodesic parallel coordinates $(x,y)$ (\ref{co}), the    metric $\dd s^2_{\kap}$ (\ref{cp}), domain of the variables,  Vessiot--Guldberg  Lie algebra  $V_\kap$ (\ref{df})  with Lie vector fields $X_{\kap,i}$ (\ref{db}),   LH algebra ${\cal H}_{\omega_\kap}$  (\ref{dg}) with Hamiltonian functions $h_{\kap,i}$ (\ref{hamfun}) (so $h_{\kap,0}=1$), and the symplectic form $\omega_\kap$ (\ref{dd}). For the sake of clarity, we drop the index $\kap=(\k_1,\k_2)$.}}
\label{table1}
\medskip
\noindent\hfill
\begin{tabular}{lll}
\hline
\\[-6pt]

$\bullet$ Sphere  ${\bf S}^2$& $\bullet$ Euclidean plane ${\bf
E}^2$
&$\bullet$ Hyperbolic  space ${\bf H}^2$  \\[2pt] 

$\mathbf S^2_{[+],+}={\rm SO(3)/SO(2)}$&$ \mathbf  S^2_{[0],+}={\rm ISO(2)/SO(2)}$&
$ \mathbf  S^2_{[-],+}={\rm SO(2,1)/SO(2)}$\\[4pt]
$\dd s^2=\cos^2 \yy \,  \dd \aa^2+ \dd\yy^2$&
$\dd s^2= \dd \aa^2+ \dd\yy^2$&
$\dd s^2=\cosh^2 \yy \, \dd \aa^2+ \dd\yy^2$\\[2pt] 
$x\in (-\pi,\pi],\  y\in (-\tfrac \pi 2, \tfrac \pi 2] $&
$x\in\mathbb R,\ y\in \mathbb R$&
$x\in\mathbb R,\ y\in \mathbb R$
\\[4pt] 

$V\simeq   {\mathfrak{so}}(3)$ & $V\simeq   {\mathfrak{iso}}(2) \simeq {\mathfrak{so}}(2) \ltimes \mathbb{R}^2   $ & $V\simeq   {\mathfrak{so}}(2,1) $\\[2pt] 
$  X_1={\partial}_x  $ & $  X_1={\partial}_x   $ & $ X_1={\partial}_x   $\\[2pt] 
$ X_{2}=\sin x \tan y\, {\partial_x} +\cos x\,{\partial_y} $ & $ X_{2} = {\partial_y} $ & $ X_{2}=-\sinh x \tanh y\, {\partial_x} +\cosh x\,{\partial_y} $\\[2pt] 
$ X_{3}=\cos x \tan y \, {\partial}_x  - \sin x\, {\partial}_y  $ & $ X_{3} = y \, {\partial}_x  -   x\, {\partial}_y   $ & $ X_{3}=\cosh x \tanh y \, {\partial}_x  - \sinh x \,{\partial}_y  $
\\[4pt] 
 
${\cal H}_\omega\simeq  \overline {\mathfrak{so}}(3) \simeq  {\mathfrak{so}}(3) \oplus \mathbb R$ & ${\cal H}_\omega\simeq   {\overline{\mathfrak{iso}}}(2) = \overline{{\mathfrak{so}}(2) \ltimes \mathbb{R}^2   }$ &${\cal H}_\omega\simeq  \overline {\mathfrak{so}}(2,1) \simeq  {\mathfrak{so}}(2,1) \oplus \mathbb R$ \\[2pt] 
$ h_{1}=\sin y  $ & $  h_{1}=  y $ & $  h_{1}=\sinh y $\\[2pt] 
$  h_{2}=-\sin x\cos y $ & $ h_{2}=-  x  $ & $  h_{2}=-\sinh x\cosh y$\\[2pt] 
 $ h_{3}= 1-\cos x\cos y   $ & $h_{3} =\tfrac 12 ( x^2 + y^2 ) $ & $ h_{3}= \cosh x\cosh y-1  $ \\[2pt] 
 $ \area=\cos \yy \,\dd \aa\wedge\dd\yy$&
$ \area=  \dd \aa\wedge\dd\yy$&
$ \area=\cosh \yy\,  \dd \aa\wedge\dd\yy$\\[6pt]

$\bullet$ Oscillating NH  space ${\bf NH}_+^{1+1}$ & $\bullet$ Galilean plane ${\bf
G}^{1+1}$
&$\bullet$ Expanding NH  space ${\bf NH}_-^{1+1}$  \\[1pt] 
(Co-Euclidean space) &  
& (Co-Minkowskian space)  \\[2pt] 

$\mathbf S^2_{[+],0}={\rm ISO(2)/\mathbb R}$&$ \mathbf  S^2_{[0],0}={\rm IISO(1)/\mathbb R}$&
$\mathbf  S^2_{[-],0}={\rm ISO(1,1)/\mathbb R}$\\[4pt]
$  {\rm d} s^2= {\rm d} x^2 ,\    {\rm d}{s'}^2= {\rm d} y^2 \  {\rm{on}} \  x= \, {\rm cte} $&
$  {\rm d} s^2= {\rm d} x^2 ,\    {\rm d}{s'}^2= {\rm d} y^2 \  {\rm{on}} \  x= \, {\rm cte} $&
$  {\rm d} s^2= {\rm d} x^2 ,\    {\rm d}{s'}^2= {\rm d} y^2 \  {\rm{on}} \  x= \, {\rm cte} $\\[2pt] 
$x\in (-\pi,\pi],\  y\in \mathbb  R$&
$x\in\mathbb R,\ y\in \mathbb R$&
$x\in\mathbb R,\ y\in \mathbb R$
\\[4pt] 

$V\simeq   {\mathfrak{iso}}(2)  \simeq {\mathfrak{so}}(2) \ltimes \mathbb{R}^2$ & $V\simeq   {\mathfrak{iiso}}(1) \simeq \mathbb R \ltimes \mathbb{R}^2   $ & $V\simeq   {\mathfrak{iso}}(1,1)  \simeq {\mathfrak{so}}(1,1) \ltimes \mathbb{R}^2 $\\[2pt] 
$  X_1={\partial}_x  $ & $  X_1={\partial}_x   $ & $ X_1={\partial}_x   $\\[2pt] 
$ X_{2}= \cos x\,{\partial_y} $ & $ X_{2} = {\partial_y} $ & $ X_{2}=  \cosh x\,{\partial_y} $\\[2pt] 
$ X_{3}=   - \sin x\, {\partial}_y  $ & $ X_{3} =    -   x\, {\partial}_y   $ & $ X_{3}=\   - \sinh x \,{\partial}_y  $
\\[4pt] 
 
${\cal H}_\omega\simeq   {\overline{\mathfrak{iso}}}(2) = \overline{{\mathfrak{so}}(2) \ltimes \mathbb{R}^2   }$& ${\cal H}_\omega\simeq   {\overline{\mathfrak{iiso}}}(1) = \overline{ \mathbb R  \ltimes \mathbb{R}^2   }$ &${\cal H}_\omega\simeq   {\overline{\mathfrak{iso}}}(1,1) = \overline{{\mathfrak{so}}(1,1) \ltimes \mathbb{R}^2   }$ \\[2pt] 
$ h_{1}=  y  $ & $  h_{1}=  y $ & $  h_{1}=  y $\\[2pt] 
$  h_{2}=-\sin x  $ & $ h_{2}=-  x  $ & $  h_{2}=-\sinh x $\\[2pt] 
 $ h_{3}= 1-\cos x    $ & $h_{3} =\tfrac 12 x^2$ & $ h_{3}= \cosh x-1  $ \\[2pt] 
 $ \area= \dd \aa\wedge\dd\yy$&
$ \area=  \dd \aa\wedge\dd\yy$&
$ \area= \dd \aa\wedge\dd\yy$\\[6pt]

$\bullet$ Anti-de Sitter space ${\bf AdS}^{1+1}$ & $\bullet$ Minkowskian plane ${\bf M}^{1+1}$
&$\bullet$ De Sitter  space ${\bf dS}^{1+1}$ \\[1pt] 
(Co-hyperbolic space) &  &
(Doubly hyperbolic space)  \\[2pt] 

$\mathbf S^2_{[+],-}={\rm SO(2,1)/SO(1,1)}$&$\mathbf  S^2_{[0],-}={\rm ISO(1,1)/SO(1,1)}$&
$\mathbf  S^2_{[-],-}={\rm SO(2,1)/SO(1,1)}$\\[4pt]
$\dd s^2=\cosh^2 \yy \,  \dd \aa^2- \dd\yy^2$&
$\dd s^2= \dd \aa^2- \dd\yy^2$&
$\dd s^2=\cos^2 \yy \,\dd \aa^2-\dd\yy^2$\\[2pt] 
$x\in (-\pi,\pi],\  y\in \mathbb  R$&
$x\in\mathbb R,\ y\in \mathbb R$&
$x\in\mathbb R,\ y\in (-\pi,\pi],$
\\[4pt] 

$V\simeq   {\mathfrak{so}}(2,1)$ & $V\simeq   {\mathfrak{iso}}(1,1) \simeq {\mathfrak{so}}(1,1) \ltimes \mathbb{R}^2   $ & $V\simeq   {\mathfrak{so}}(2,1) $\\[2pt] 
$  X_1={\partial}_x  $ & $  X_1={\partial}_x   $ & $ X_1={\partial}_x   $\\[2pt] 
$ X_{2}=-\sin x \tanh y\, {\partial_x} +\cos x\,{\partial_y} $ & $ X_{2} = {\partial_y} $ & $ X_{2}=\sinh x \tan y\, {\partial_x} +\cosh x\,{\partial_y} $\\[2pt] 
$ X_{3}=-\cos x \tanh y \, {\partial}_x  - \sin x\, {\partial}_y  $ & $ X_{3} =- y \, {\partial}_x  -   x\, {\partial}_y   $ & $ X_{3}=-\cosh x \tan y \, {\partial}_x  - \sinh x \,{\partial}_y  $
\\[4pt] 
 
${\cal H}_\omega\simeq  \overline {\mathfrak{so}}(2,1) \simeq  {\mathfrak{so}}(2,1) \oplus \mathbb R$ & ${\cal H}_\omega\simeq   {\overline{\mathfrak{iso}}}(1,1) = \overline{{\mathfrak{so}}(1,1) \ltimes \mathbb{R}^2   }$ &${\cal H}_\omega\simeq  \overline {\mathfrak{so}}(2,1) \simeq  {\overline{{\mathfrak{so}}(2,1) \oplus \mathbb R}}$ \\[2pt] 
$ h_{1}=\sinh y  $ & $  h_{1}=  y $ & $  h_{1}=\sin y $\\[2pt] 
$  h_{2}=-\sin x\cosh y $ & $ h_{2}=-  x  $ & $  h_{2}=-\sinh x\cos y$\\[2pt] 
 $ h_{3}= 1-\cos x\cosh y   $ & $h_{3} =\tfrac 12 ( x^2 - y^2 ) $ & $ h_{3}= \cosh x\cos y-1  $ \\[2pt] 
 $ \area=\cosh \yy \,\dd \aa\wedge\dd\yy$&
$ \area=  \dd \aa\wedge\dd\yy$&
$ \area=\cos \yy\,  \dd \aa\wedge\dd\yy$\\[6pt]

\hline
\end{tabular}\hfill}
\end{table}

\section{Constants of motion and superposition rules}

This section deals with the computations of the constants of motion for the LH system $X_\kap$ (\ref{dc}), which will further allow us to deduce the corresponding superposition rules by applying the Poisson coalgebra approach~\cite{BHLS,BCHLS13Ham}.


\subsection{Constants of motion}

Likewise in section 2.1, the space $S\left( \overline{ {\mathfrak{so}}}_\kap(3) \right)$  stands for the symmetric algebra  of the extended CK Lie algebra  $ \overline{ {\mathfrak{so}}}_\kap(3) $. The symmetric algebra is naturally a Poisson algebra. Consider a basis  $\{ v_1,v_2,v_3,v_0\}$  of  $ \overline{ {\mathfrak{so}}}_\kap(3) $   satisfying the commutation relations (\ref{dg}). Then,  the element
\be
C_\kap:=v_{3}v_{0}-\frac{1}{2}\bigl(\k_2 v_{1}^{2}+v_{2}^{2}+\k_1 v_3^2\bigr) 
\label{fb}
\ee
Poisson commutes with all $v_a$,  i.e. it is a  second-order  Casimir  (invariant)  of $S\left( \overline{ {\mathfrak{so}}}_\kap(3) \right)$.
Next, we consider the non-deformed   coproduct map ${\Delta} :S\left( \overline{ {\mathfrak{so}}}_\kap(3) \right) \rightarrow
S\left( \overline{ {\mathfrak{so}}}_\kap(3) \right) \otimes S\left( \overline{ {\mathfrak{so}}}_\kap(3) \right)$ given by  (\ref{aj}) along with the 
Poisson algebra morphisms $D: S\left( \overline{ {\mathfrak{so}}}_\kap(3) \right)   \rightarrow C^\infty({\mathbf S}^2_{[\k_1],\k_2}) $ and  $D^{(2)} :   S\left(   \overline{ {\mathfrak{so}}}_\kap(3) \right)\otimes S\left( \overline{ {\mathfrak{so}}}_\kap(3) \right)\rightarrow C^\infty(  {\mathbf S}^2_{[\k_1],\k_2}   )\otimes C^\infty( {\mathbf S}^2_{[\k_1],\k_2})$ defined, similarly to   (\ref{ak}),  by
$$
\begin{gathered}
D( v_a):= h_{\kap,a}(x_1,y_1), \quad
 D^{(2)} \left(v_a\otimes 1\right):= h_{\kap,a}(x_1,y_1),\quad D^{(2)} \left(1\otimes v_a\right):=h_{\kap,a}(x_2,y_2),   
\label{fc}\nonumber
\end{gathered}
$$
$(a=0,1,2,3)$ where $h_{\kap,a}$ are now the Hamiltonians  functions (\ref{hamfun}). This gives rise to two $t$-independent constants of motion for the system $X_\kap$ (\ref{dc}) of the form
\be
  F_\kap:= D(C_\kap),\qquad F_\kap^{(2)}:=  D^{(2)} \left( {\Delta}(C_\kap) \right) .
\label{fd}\nonumber
\ee
The former turns out to be  trivial, $  F_\kap=0$, meanwhile the latter can be written as
\bea
&&
F_\kap^{(2)} = 
\frac{1}{\kk_1}\bigl(1-\Ck_{\kk_1}(x_1-x_2)\Ck_{\kk_1\kk_2}(y_1)\Ck_{\kk_1\kk_2}(y_2)-\kk_1\kk_2
\Sk_{\kk_1\kk_2}(y_1)\Sk_{\kk_1\kk_2}(y_2) \bigr) \nonumber\\[2pt]
&&\qquad \, =  \Vk_{\kk_1}(x_1-x_2)\Ck_{\kk_1\kk_2}(y_1)\Ck_{\kk_1\kk_2}(y_2)+ \kk_2
\Vk_{\kk_1\kk_2}(y_1-y_2)  ,
\label{fe}
\eea
where we have used the relations
 \be
 \Ck_\k(u\pm v)= \Ck_\k(u) \Ck_\k(v)\mp \k  \Sk_\k(u) \Sk_\k(v),\quad \Sk_\k(u\pm v)= \Sk_\k(u) \Ck_\k(v)\pm   \Ck_\k(u) \Sk_\k(v).
\label{zc}
\ee
 Note that the second expression for $F_\kap^{(2)} $ admits the direct flat contraction $\k_1=0$; explicitly, since $\Vk_{0}(u) =u^2/2$, then
 $$
 F_{\k_1=0,\k_2}^{(2)} = \frac{1}{2}\left[
(x_{1}-x_{2})^{2}+\k_2 (y_{1}-y_{2})^{2} \right], 
 $$
 so that for $\k_2=+1$   we recover the Euclidean constant of motion (\ref{am}).

We stress that, in fact,  this constant of motion corresponds to the geodesic  distance $s_1$ between two points $(x_1,y_1)$ and $(x_2,y_2)$ on the space  $\mathbf  S^2_{[\kk_1],\kk_2}$, which is given by~\cite{HS02}
\be
\Ck_{\kk_1}( s_1)= \Ck_{\kk_1}(x_1-x_2)\Ck_{\kk_1\kk_2}(y_1)\Ck_{\kk_1\kk_2}(y_2)+\kk_1\kk_2
\Sk_{\kk_1\kk_2}(y_1)\Sk_{\kk_1\kk_2}(y_2) .
\label{ff}
\ee

Recall that $F_\kap^{(2)}$ is a $t$-independent constant of motion for the diagonal prolongation $\widetilde{X}_{\kap}$ of $X_\kap$ to the manifold 
$\mathbf S^2_{[\kk_1],\kk_2}\times\, \mathbf  S^2_{[\kk_1],\kk_2}$ (cf.~\cite{Dissertations}); namely, if 
$X_\kap=X(x,y) \partial_x+Y(x,y) \partial_y$, then
$$
\widetilde{X}_\kap=X(x_{1},y_{1})\frac{\partial}{\partial x_{1}}+Y(x_{1},y_{1})\frac{\partial}{\partial y_{1}}+
X(x_{2},y_{2})\frac{\partial}{\partial x_{2}}+Y(x_{2},y_{2})\frac{\partial}{\partial y_{2}},
$$
where $((x_{1},y_{1}),(x_{2},y_{2}))\in \mathbf  S^2_{[\kk_1],\kk_2}\times\, \mathbf  S^2_{[\kk_1],\kk_2}$. Moreover, the function $F_\kap^{(2)}$ gives rise to two other constants of motion  through the  permutation $S_{ij}$ of the variables $(x_i,y_i)\leftrightarrow (x_j,y_j)$; these are
\be
F^{(2)}_{\kap,13}=S_{13}(F_\kap^{(2)}),\qquad F^{(2)}_{\kap,23}=S_{23}(F_\kap^{(2)}) .
\label{fg}
\ee
Since prolongations are invariant under permutations, the functions $F^{(2)}_{\kap,ij}$ are also $t$-independent constants of motion for the diagonal prolongations
$\widetilde{X}_\kap$ to $\mathbf S^2_{[\kk_1],\kk_2}\times \mathbf S^2_{[\kk_1],\kk_2}$.

By taking into account the expressions  (\ref{fe}), (\ref{ff}),  and (\ref{fg}), we can write the above three constants of motion in the form 
\bea
&& F_\kap^{(2)}=\frac{1}{\kk_1}\bigl(1-\Ck_{\kk_1}(x_1-x_2)\Ck_{\kk_1\kk_2}(y_1)\Ck_{\kk_1\kk_2}(y_2)-\kk_1\kk_2
\Sk_{\kk_1\kk_2}(y_1)\Sk_{\kk_1\kk_2}(y_2) \bigr) \nonumber\\
&&\qquad\,   =\frac{1}{\kk_1} \bigr(1-\Ck_{\kk_1}(s_1) \bigr)=\Vk_{\kk_1}(s_1) , \nonumber\\
&&F^{(2)}_{\kap,23}=\frac{1}{\kk_1} \bigl(1-\Ck_{\kk_1}(x_1-x_3)\Ck_{\kk_1\kk_2}(y_1)\Ck_{\kk_1\kk_2}(y_3)-\kk_1\kk_2
\Sk_{\kk_1\kk_2}(y_1)\Sk_{\kk_1\kk_2}(y_3) \bigr)\nonumber\\
&&\qquad\    =\frac{1}{\kk_1} \bigr(1-\Ck_{\kk_1}(s_2) \bigr) =\Vk_{\kk_1}(s_2), \nonumber\\
&&F^{(2)}_{\kap,13}=\frac{1}{\kk_1} \bigl(1-\Ck_{\kk_1}(x_3-x_2)\Ck_{\kk_1\kk_2}(y_3)\Ck_{\kk_1\kk_2}(y_2)-\kk_1\kk_2
\Sk_{\kk_1\kk_2}(y_3)\Sk_{\kk_1\kk_2}(y_2) \bigr)\nonumber\\
&&\qquad\   =\frac{1}{\kk_1} \bigr(1-\Ck_{\kk_1}(s_3) \bigr)=\Vk_{\kk_1}(s_3),
\label{constofmot}
\eea
where $s_1,s_2, s_3$ are three positive  real constants.


\subsection{Superposition rules}

Since $\partial(F_\kap^{(2)},F^{(2)}_{\kap,23})/\partial{(x_1,y_1)}\neq 0$, both constants of motion (\ref{constofmot}) are   functionally  independent functions. This, in turn, means  that one can express the general solution $(x_1(t),y_1(t))$  of the LH system $X_\kap$  (\ref{dc}) in terms of two different particular solutions
$(x_2(t),y_2(t))$,  $(x_3(t),y_3(t))$, and  the two constants $s_1$, $s_2$.  Therefore, one may start  with such explicit expressions   (\ref{constofmot}) and try to compute the superposition rules.   Nevertheless, such a `direct' procedure is frequently cumbersome  and non-trivial. By contrast, we shall be able to  obtain a closed  analytical form for the   superposition rules by applying a geometric approach based on the  trigonometry of the CK spaces ${\mathbf S}^2_{[\k_1],\k_2}$ (\ref{cc}).  All the trigonometric relations as well as  generalized theorems used in our procedure can be found in~\cite{HOS00}.

With this aim, we set the points  $Q_1:=(x_1,y_1)$, $Q_2:=(x_2,y_2)$, and
$Q_3:=(x_3,y_3)$ in $\mathbf S^2_{[\kk_1],\kk_2}$ forming  a triangle $\triangle Q_1Q_2Q_3$. Its sides are geodesics such that  the positive constants  $s_1$, $s_2$ and $s_3$ appearing in (\ref{constofmot})  are, in this order,  the geodesic distances   $\overline{Q_1Q_2}$,
$\overline{Q_1Q_3}$ and $\overline{Q_3Q_2}$, so fulfilling (\ref{ff}), and $\ados$ is the angle between the geodesics $ {Q_1Q_2}$ and $ {Q_3Q_2}$; this is depicted in figure~\ref{figure2}.

Consider   the   orthogonal triangle $\triangle Q_1Q_2Q_{12}$ where  $Q_{12}=(x_1,y_2)$, such that    the geodesics  $ {Q_{12}Q_2}$  and  $ {Q_1Q_{12}}$  are orthogonal at $Q_{12}$ and with inner angle $\ados+\auno$ at $Q_2$, as shown in    figure~\ref{figure2}. 
The cosine and sine theorems for this triangle, with geodesic distances $\overline{Q_1Q_{12}}=y_1-y_2>0$ and $\overline{Q_{12}Q_2}=x_1-x_2>0$,  read
\be
\Ck_{\k_1}(s_1) =\Ck_{\k_1}(x_1-x_2)\Ck_{\k_1\k_2}(y_1-y_2),\qquad \Sk_{\kk_1\kk_2}(y_1-y_2)=\Sk_{\kk_1}(s_1)\Sk_{\kk_2}(\ados+\auno).
\label{ga}\nonumber
\ee
From these, we find that
\be
\Tk_{\k_1}( x_1-x_2)=\Tk_{\k_1}(s_1)\Ck_{\k_2}(  \ados+\auno).
\label{gb} \nonumber
\ee
After using the formulas (\ref{zc}), we arrive at
\be
\begin{aligned}
&\Tk_{\kk_1}(x_1-x_2)=\Tk_{\kk_1}(s_1)\left[\Ck_{\kk_2}(\ados)\Ck_{\kk_2}(\auno) -\kk_2
\Sk_{\kk_2}(\ados) \Sk_{\kk_2}(\auno)\right],\\
&\Sk_{\kk_1\kk_2}(y_1-y_2)=\Sk_{\kk_1}(s_1)\left[\Sk_{\kk_2}(\ados)
\Ck_{\kk_2}(\auno) + \Sk_{\kk_2}(\auno)\Ck_{\kk_2}(\ados)\right].
\label{suprule}
\end{aligned}
\ee
Therefore, we need to express $\Ck_{\kk_2}(\ados)$, $\Sk_{\kk_2}(\ados)$, $\Ck_{\kk_2}(\auno)$ and $\Sk_{\kk_2}(\auno)$  in terms of $(x_2(t),y_2(t))$, $(x_3(t),y_3(t))$  and the positive constants $s_1$, $s_2$, $s_3$.


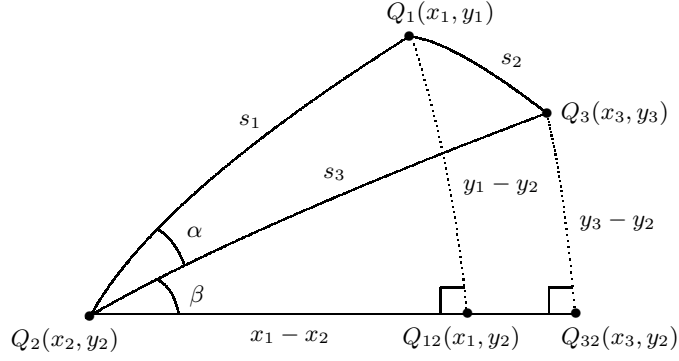
\begin{figure}[t]
\begin{center}
\begin{picture}(220,140)
\footnotesize{
\put(15,15){\makebox(0,0){$Q_2(x_2,y_2)$}}
\put(22,22){$\bullet$}
\put(157,139){\makebox(0,0){$Q_1(x_1,y_1)$}}
 \put(223,101){\makebox(0,0){$Q_3(x_3,y_3)$}}
 \put(224,60){\makebox(0,0){$y_3-y_2$}}
 \put(180,74){\makebox(0,0){$y_1-y_2$}}
\put(195,99){$\bullet$}
  \put(85,15){$x_1-x_2$}
\put(85,98){\makebox(0,0){$s_1$}}
\put(117,78){\makebox(0,0){$s_3$}}
\put(184,121){\makebox(0,0){$s_2$}}
 \put(23,25){\line(1,0){185}}  
\put(143,128){$\bullet$}
\put(165,23){$\bullet$}  
\put(65,32){\makebox(0,0){$\auno$}}
\put(64,56){\makebox(0,0){$\ados$}}
\qbezier(25,25)(50,70)(145,130)  
 \put(-2, 11){\qbezier(25,12.5)(80,45)(199,90)}  
  \put(132,133){\qbezier(14,-3)(30,-05)(64,-31)} 
 \put(78,0){\qbezier[40](130,25)(125,80)(119,101)}  
\put(206,23){$\bullet$}  
\put(225,16){\makebox(0,0){$Q_{32}(x_3,y_2)$}}
\put(165,16){\makebox(0,0){$Q_{12}(x_1,y_2)$}}
 \put(37,0){\qbezier[50](130,25)(125,80)(110,127)}  
\put(198,25){\line(0,1){10}}
\put(198,35){\line(1,0){9}}
  \put(157,25){\line(0,1){10}}
\put(157,35){\line(1,0){9}}
 \put(12,0){ \qbezier(46,25)(44,35)(38,38) }  
 \put(12,19){ \qbezier(48,25)(44,35)(38,38) }  
}
\end{picture}
\end{center}
\noindent
\\[-50pt]
\caption{\footnotesize Triangles and geodesic distances involved in the derivation of the superposition rules for the LH system $X_\kap$ (\ref{dc}) on the CK space ${\mathbf S}^2_{[\k_1],\k_2}$ (\ref{cc}).}
\label{figure2}
\end{figure}


Firstly, as above, if we now take the  orthogonal triangle $\triangle Q_2Q_3Q_{32}$ where $Q_{32}=(x_3,y_2)$ with the geodesics    $ {Q_{32}Q_2}$  and  $ {Q_3Q_{32}}$   being orthogonal at $Q_{32}$ and with inner angle $\auno$ at $Q_2$, we can write
\be
\Ck_{\k_2}(\auno)= \frac{\Tk_{\k_1}(x_3-x_2)}{\Tk_{\k_1}(s_3)},\qquad  \Sk_{\k_2}(\auno)= \frac{\Sk_{\k_1\k_2}(y_3-y_2)}{\Sk_{\k_1}(s_3)}, 
\label{gc}
\ee
where we have made use of the  geodesic distances $\overline{Q_{32}Q_{2}}=x_3-x_2>0$ and $\overline{Q_3Q_{32}}=y_3-y_2>0$.

Secondly, in order to get $\Ck_{\kk_2}(\ados)$ and $\Sk_{\kk_2}(\ados)$, we consider the `initial'  triangle $\triangle Q_1Q_2Q_3$. By one hand, the cosine theorem gives
\be
\Ck_{\kk_2}(\ados)=\frac{\Ck_{\kk_1}(s_2)- \Ck_{\kk_1}(s_1)\Ck_{\kk_1}(s_3)}{ \kk_1\Sk_{\kk_1}(s_1)\Sk_{\kk_1}(s_3)} .
\label{gd}
\ee
  On the other hand,   $\Sk_{\kk_2}(\ados)$ can be written in terms of  the area $A$ of this triangle through the generalized Cagnoli's theorem
  \be
\Sk_{\kk_2}(\ados)= \frac{4\Ck_{\kk_1}(\frac{s_1}{2})\Ck_{\kk_1}(\frac{s_2}{2})\Ck_{\kk_1}(\frac{s_3}{2})
\Sk_{\kk_1^2\kk_2}(\frac{A}{2})}{\Sk_{\k_1}(s_1)\Sk_{\kk_1}(s_3)}.
\label{ge}\ee

Consequently, by substituting (\ref{gc}), (\ref{gd}) and (\ref{ge}) in (\ref{suprule}), we obtain that
\bea
&&\!\!\!\!\!\! 
\Tk_{\kk_1}(x_1-x_2)=\Tk_{\kk_1}(x_3-x_2)\, \frac{\Ck_{\kk_1}(s_2)-\Ck_{\kk_1}(s_1)\Ck_{\kk_1}(s_3)}{\kk_1\Ck_{\kk_1}(s_1)
\Sk_{\kk_1}(s_3)\Tk_{\kk_1}(s_3)}\nonumber\\[2pt]
&&\qquad\qquad \qquad\quad - 4\kk_2\Sk_{\kk_1\kk_2}(y_3-y_2) \,  \frac{ \Ck_{\kk_1}(\frac{s_1}{2})\Ck_{\kk_1}(\frac{s_2}{2})\Ck_{\kk_1}(\frac{s_3}{2})
\Sk_{\kk_1^2\kk_2}(\frac{A}{2}) }{\Ck_{\kk_1}(s_1)\Sk_{\kk_1}^2(s_3)}, \nonumber\\[2pt]
&&\!\!\!\!\!\! 
\Sk_{\kk_1\k_2}(y_1-y_2)=\Sk_{\kk_1\k_2}(y_3-y_2)\, \frac{\Ck_{\kk_1}(s_2)-\Ck_{\kk_1}(s_1)\Ck_{\kk_1}(s_3)}{\kk_1
\Sk_{\kk_1}^2(s_3)}\nonumber\\[2pt]
&&\qquad\qquad \qquad\quad + 4\Tk_{\kk_1 }(x_3-x_2)\,  \frac{ \Ck_{\kk_1}(\frac{s_1}{2})\Ck_{\kk_1}(\frac{s_2}{2})\Ck_{\kk_1}(\frac{s_3}{2})
\Sk_{\kk_1^2\kk_2}(\frac{A}{2})   }{\Sk_{\kk_1}(s_3)\Tk_{\kk_1}(s_3)}.
\label{gf}
\eea
These relations can  further be    written in different ways by considering the expressions for the area $A$ presented in~\cite{HOS00}. For instance,
for the six spaces with $\k_2\ne 0$ (so precluding the three Newtonian spaces),  there exists a  generalized {\it L'Huillier formula} $A=A(s_1,s_2,s_3)$, which is the curved counterpart of the Heron--Archimedes area formula (\ref{ap}) for the Euclidean plane with $(\k_1,\k_2)=(0,+1)$, given by
\bea
&& \Tk^2_{\kk_1^2\kk_2}\left(\frac{A}{4}\right)=\frac 1 {\k_2}   \Tk_{\kk_1}\left( \frac{p}{2} \right) \Tk_{\kk_1}\left( \frac{p-s_1}{2} \right)   \Tk_{\kk_1}\left( \frac{p-s_2}{2} \right) \Tk_{\kk_1}\left( \frac{p-s_3}{2} \right)    ,\qquad \k_2\ne 0, \nonumber\\[2pt]
&&  p=\frac 12 (s_1+s_2+s_3)  ,
\label{Heron}\nonumber
\eea
such that  $p$ is one-half the sum of the three  geodesic sides of the triangle $\triangle Q_1Q_2Q_3$. In particular, for the flat Euclidean $(\k_2>0)$ and Minkowskian  $(\k_2<0)$  spaces, this expression reduces to
$$
 A^2= \frac 1 {\k_2}\, p (p-s_1)(p-s_2)(p-s_3)= \frac 1{16 \k_2} \left[ {2(s_1^2s_2^2+s_1^2s_3^2+s_2^2s_3^2)-(s_1^4+s_2^4+s_3^4)} \right],
$$
which is just (\ref{ap}) for $\k_2=+1$ and $s_i:= k_i$.

We stress that there exists a second solution for the superposition rules (see the Euclidean case (\ref{ap})), say $(x^-_1,y^-_1)$, which corresponds to change the sign of the last term in both relations (\ref{gf}). This can be proven in a similar way by considering another configuration for the  triangles.

We summarize the results of this section in the following statement.

 \medskip
\noindent
{\bf Theorem 1.}  {\em Let $X_\kap$ be the LH system (\ref{dc}) defined on the CK space $\mathbf S^2_{[\kk_1],\kk_2}$, 
with vector fields (\ref{db}), Hamiltonian functions (\ref{hamfun}) and symplectic form (\ref{dd}). Then:\\
\noindent
(i) The functions (\ref{constofmot}) are three $t$-independent constants of motion for the  diagonal prolongation $\widetilde{X}_\kap$ to the manifold 
$\mathbf S^2_{[\kk_1],\kk_2}\times\, \mathbf  S^2_{[\kk_1],\kk_2}\times\, \mathbf  S^2_{[\kk_1],\kk_2}$, such that any pair among them is formed by  two functionally independent functions.\\
(ii) The     general solution $(x_1(t),y_1(t))$ of $X_\kap$ in terms of two different particular solutions $(x_2(t),y_2(t))$ and $(x_3(t),y_3(t))$  can be written as\bea
&&\!\!\!\!\!\! 
\Tk_{\kk_1}(x_1^\pm -x_2)=\Tk_{\kk_1}(x_3-x_2)\, \frac{\Ck_{\kk_1}(s_2)-\Ck_{\kk_1}(s_1)\Ck_{\kk_1}(s_3)}{\kk_1\Ck_{\kk_1}(s_1)
\Sk_{\kk_1}(s_3)\Tk_{\kk_1}(s_3)}\nonumber\\[2pt]
&&\qquad\qquad \qquad\quad \mp 4\kk_2\Sk_{\kk_1\kk_2}(y_3-y_2) \,  \frac{ \Ck_{\kk_1}(\frac{s_1}{2})\Ck_{\kk_1}(\frac{s_2}{2})\Ck_{\kk_1}(\frac{s_3}{2})
\Sk_{\kk_1^2\kk_2}(\frac{A}{2}) }{\Ck_{\kk_1}(s_1)\Sk_{\kk_1}^2(s_3)}, \nonumber\\[2pt]
&&\!\!\!\!\!\! 
\Sk_{\kk_1\k_2}(y_1^\pm -y_2)=\Sk_{\kk_1\k_2}(y_3-y_2)\, \frac{\Ck_{\kk_1}(s_2)-\Ck_{\kk_1}(s_1)\Ck_{\kk_1}(s_3)}{\kk_1
\Sk_{\kk_1}^2(s_3)}\nonumber\\[2pt]
&&\qquad\qquad \qquad\quad \pm 4\Tk_{\kk_1 }(x_3-x_2)\,  \frac{ \Ck_{\kk_1}(\frac{s_1}{2})\Ck_{\kk_1}(\frac{s_2}{2})\Ck_{\kk_1}(\frac{s_3}{2})
\Sk_{\kk_1^2\kk_2}(\frac{A}{2})   }{\Sk_{\kk_1}(s_3)\Tk_{\kk_1}(s_3)},
\label{gi}
\eea
where $s_1,s_2,s_3$, and $A$ are   positive constants, such that the latter is just the area of triangle formed by the three solutions considered as points in  $\mathbf S^2_{[\kk_1],\kk_2}$, meanwhile the former ones are its geodesic sides.
}


\subsection{Discussion}

We illustrate the above results by writing the Casimir (\ref{fb}), the constant of motion   $F_\kap^{(2)}$  (\ref{constofmot}), and the superposition rules (\ref{gi}) in table~\ref{table2} for each of the nine CK spaces  ${\mathbf
S}^2_{[\k_1],\k_2}$ (\ref{cc})  with the same structure of table~\ref{table1}. We remark that tables~\ref{table1} and \ref{table2} comprise the main results of the paper.
When both tables are read by rows, one finds, in this order, the  three  classical Riemannian spaces  of constant  curvature $\k_1$ with $\k_2>0$, the   semi-Riemannian spaces or Newtonian spacetimes  of constant  curvature $\k_1$ with $\k_2=0$ ($c\to \infty$), and the   pseudo-Riemannian spaces or Lorentzian spacetimes of constant  curvature $\k_1=-\Lambda$ with $\k_2=-1/c^2<0$. When these are read by columns, one finds three spaces of positive, zero (flat) and negative constant curvature, correspondingly, but with different metric signature ${\rm diag}(+1,\k_2)$.

 Clearly, all the Euclidean results previously obtained in~\cite{{BHLS}}, and here summarized in section 2,  are recovered for  ${\mathbf
S}^2_{[0],+}$  such that the three constants $k_i$ $(i=1,2,3)$ coincide with the geodesic distances $s_i$.
We recall that  for the three flat spaces with $\k_1=0$ (middle column of the tables), ${\mathbf
S}^2_{[0],\k_2}$,  the contraction of the constant of motion   $F_\kap^{(2)}$  comes out directly from the  expression written in terms 
of  $\k$-versed sines in (\ref{fe}). Similarly, the contraction of the superposition rules (\ref{gi}) can be obtained  by taking into account that the factor
$$
\frac{\Ck_{\kk_1}(s_2)-\Ck_{\kk_1}(s_1)\Ck_{\kk_1}(s_3)}{\kk_1 }=\!\Vk_{\kk_1}(s_1)\!+\!\Vk_{\kk_1}(s_3)\!-\!\Vk_{\kk_1}(s_2)\!-\!\k_1 \Vk_{\kk_1}(s_1) \Vk_{\kk_1}(s_3)\ \ 
{\mapsto}\ \   \frac 12 (s_1^2+s_3^2-s_2^2),
$$
when $\k_1=0$, so avoiding to take power series in the curvature.

An important fact concerns the three semi-Riemannian   or Newtonian spaces with $\k_2=0$  (middle row of the tables), ${\mathbf
S}^2_{[\k_1],0}$. The constants of motion   (\ref{constofmot}) only includes the variables  $x_i$ $(i=1,2,3)$, that is, 
$$
 F_{\k_1,\k_2=0}^{(2)}=\frac{1}{\kk_1}\bigl(1-\Ck_{\kk_1}(x_1-x_2) \bigr)= \frac{1}{\kk_1}\bigl(1-\Ck_{\kk_1}(s_1)\bigr).
 $$
Therefore,  strictly speaking, only   a part of their superposition rules can be derived from them, which corresponds to the first relation of (\ref{gi}):
$$
\Tk_{\kk_1}(x_1^\pm -x_2)=\Tk_{\kk_1}(x_3-x_2)\, \frac{\Ck_{\kk_1}(s_2)-\Ck_{\kk_1}(s_1)\Ck_{\kk_1}(s_3)}{\kk_1\Ck_{\kk_1}(s_1)
\Sk_{\kk_1}(s_3)\Tk_{\kk_1}(s_3)}.
$$
 Nevertheless, we stress that the complete superposition rules (\ref{gi}) also apply for these spaces in such a manner that the `missing' part, containing the variables $y_i$, is consistently obtained through the contraction procedure. This corresponds to the second expression in  (\ref{gi}), namely
\be
 y_1^\pm  -y_2 = (y_3-y_2)\, \frac{\Ck_{\kk_1}(s_2)-\Ck_{\kk_1}(s_1)\Ck_{\kk_1}(s_3)}{\kk_1
\Sk_{\kk_1}^2(s_3)}   \pm 2 \Tk_{\kk_1 }(x_3-x_2)\,  \frac{ \Ck_{\kk_1}(\frac{s_1}{2})\Ck_{\kk_1}(\frac{s_2}{2})\Ck_{\kk_1}(\frac{s_3}{2})
  {A}  }{\Sk_{\kk_1}(s_3)\Tk_{\kk_1}(s_3)} .
\label{gpp}
\ee
  Let us explain this point   from a trigonometry procedure. Consider the triangles of figure~\ref{figure2} that represent the solution $(x_1^+,y_1^+)$ of the superposition rules. Trigonometry on ${\mathbf
S}^2_{[\k_1],0}$ gives the relations
$$
s_1=x_1-x_2,\qquad s_2=x_3-x_1,\qquad s_3=x_3-x_2,\qquad s_3=s_1+s_2 ,
$$
 so that
\be
 \Sk_{\k_1}(s_1)=\frac{\Ck_{\k_1}(s_2)-\Ck_{\k_1}(s_1)\Ck_{\k_1}(s_3)}{\k_1 \Sk_{\k_1}(s_3)},\qquad \Tk_{\k_1}(s_3)=\Tk_{\k_1}(x_3-x_2) .
\label{gp}
\ee
 The  sine theorem  on the orthogonal triangles $\triangle  Q_1Q_2Q_{12}$ and $\triangle  Q_2Q_3Q_{32}$ reads
 \be
 y_1-y_2= \Sk_{\k_1}(s_1) (\alpha+\beta),\qquad y_3-y_2= \Sk_{\k_1}(s_3) \,\beta .
\label{gq}
\ee

\begin{landscape}

\begin{table}[htbp]
  \vskip-1cm
{\footnotesize
 \noindent
\caption{{\small For each LH system $X_\kap$  (\ref{dc}) on the CK  space ${\mathbf
S}^2_{[\k_1],\k_2}$ (\ref{cc}) it is displayed, according to the `normalized'  values of the contraction parameters  $\k_a\in\{1, 0, -1\}$,  the  Casimir $C_\kap$ (\ref{fb}), the constant of motion $F_\kap^{(2)}$  (\ref{constofmot}) and   the superposition rules (\ref{gi})  in geodesic parallel coordinates $(x,y)$ (\ref{co}).}}
\label{table2}
\medskip
\noindent\hfill
\begin{tabular}{lll}
\hline
\\[-6pt]

$\bullet$ Sphere  $\mathbf S^2_{[+],+}={\bf S}^2$& $\bullet$ Euclidean plane $\mathbf  S^2_{[0],+}={\bf
E}^2$
&$\bullet$ Hyperbolic  space $ \mathbf  S^2_{[-],+}={\bf H}^2$  \\[4pt] 

$ C=v_{3}v_{0}-\tfrac{1}{2}\bigl( v_{1}^{2}+v_{2}^{2}+ v_3^2\bigr) $&
$ C=v_{3}v_{0}-\tfrac{1}{2}\bigl( v_{1}^{2}+v_{2}^{2} \bigr) $&
$ C=v_{3}v_{0}-\tfrac{1}{2}\bigl(  v_{1}^{2}+v_{2}^{2}-v_3^2\bigr) $\\[2pt]

$F^{(2)}= 1-\cos(x_1-x_2) \cos y_1 \cos y_2-  \sin y_1 \sin y_2      $&
$F^{(2)}=\tfrac{1}{2}\left[
(x_{1}-x_{2})^{2}+(y_{1}-y_{2})^{2} \right] $&
$F^{(2)} =\cosh(x_1-x_2) \cosh  y_1 \cosh  y_2-
\sinh y_1 \sinh y_2-1 $\\[2pt]
$  \qquad\, =1- \cos s_1  $& $\qquad\, =\tfrac 12 s_1^2$  & $ \qquad\, = \cosh s_1 -1  $\\

$ \displaystyle { \tan(x_1 -x_2)=\tan(x_3-x_2) \frac{ \cos s_2-\cos  s_1 \cos  s_3}{ \cos s_1
 \sin s_3  \tan s_3} }$  &$ \displaystyle {  x_1 -x_2=(x_3-x_2) \frac{    s_1^2+    s_3^2-s_2^2 }{ 2
   s_3^2} }$  & $ \displaystyle { \tanh(x_1 -x_2)= \tanh(x_3-x_2) \frac{ \cosh  s_1 \cosh  s_3-\cosh s_2}{ \cosh s_1
 \sinh s_3  \tanh s_3} }$  \\[8pt]
$ \displaystyle { \qquad \mp 4 \sin(y_3-y_2)    \frac{ \cos(\frac{s_1}{2})\cos(\frac{s_2}{2})\cos(\frac{s_3}{2})
\sin(\frac{A}{2}) }{\cos s_1\sin^2 s_3}  }$  &$ \displaystyle { \qquad\qquad\quad \mp 2  (y_3-y_2)    \frac{  
  {A}  }{  s_3^2}  }$  & $ \displaystyle { \qquad \mp 4 \sinh(y_3-y_2)    \frac{ \cosh(\frac{s_1}{2})\cosh(\frac{s_2}{2})\cosh(\frac{s_3}{2})
\sin(\frac{A}{2}) }{\cosh s_1\sinh^2 s_3}  }$  \\[8pt]

$ \displaystyle {  \sin(y_1  -y_2)=\sin(y_3-y_2)  \frac{ \cos s_2- \cos s_1 \cos s_3 }{ 
\sin^2s_3} }$  &$ \displaystyle {   y_1 -y_2=(y_3-y_2) \frac{    s_1^2+    s_3^2-s_2^2 }{ 2
   s_3^2}   }$  & $ \displaystyle { \sinh(y_1  -y_2)=\sinh(y_3-y_2)  \frac{ \cosh s_1 \cosh s_3-\cosh s_2 }{ 
\sinh^2s_3}  }$  \\[8pt]
$ \displaystyle { \qquad \pm 4 \tan(x_3-x_2)    \frac{ \cos(\frac{s_1}{2})\cos(\frac{s_2}{2})\cos(\frac{s_3}{2})
\sin(\frac{A}{2}) }{\sin s_3\tan s_3}  }$  &$ \displaystyle { \qquad\qquad\quad \pm 2  (x_3-x_2)    \frac{  
  {A}  }{  s_3^2}  }$  & $ \displaystyle { \qquad \pm 4 \tanh(x_3-x_2)    \frac{ \cosh(\frac{s_1}{2})\cosh(\frac{s_2}{2})\cosh(\frac{s_3}{2})
\sin(\frac{A}{2}) }{\sinh s_3\tanh   s_3}  }$  \\[12pt]

$\bullet$ Oscillating NH  space $\mathbf S^2_{[+],0}={\bf NH}_+^{1+1}$ & $\bullet$ Galilean plane $\mathbf  S^2_{[0],0}={\bf
G}^{1+1}$
&$\bullet$ Expanding NH  space $\mathbf  S^2_{[-],0}={\bf NH}_-^{1+1}$  \\[4pt] 
 
$ C=v_{3}v_{0}-\tfrac{1}{2}\bigl(  v_{2}^{2}+ v_3^2\bigr) $&
$ C=v_{3}v_{0}-\tfrac{1}{2} v_{2}^{2}  $&
$ C=v_{3}v_{0}-\tfrac{1}{2}\bigl(  v_{2}^{2}-v_3^2\bigr) $\\[2pt]

$F^{(2)}= 1- \cos(x_1-x_2)   = 1-\cos s_1  $& $F^{(2)}=\tfrac{1}{2} 
(x_{1}-x_{2})^{2} =\tfrac 12 s_1^2$& $F^{(2)}= \cosh(x_1-x_2) -1  = \cosh s_1 -1 $\\[2pt]

$ \displaystyle { \tan(x_1 -x_2)=\tan(x_3-x_2) \frac{ \cos s_2-\cos  s_1 \cos  s_3}{ \cos s_1
 \sin s_3  \tan s_3} }$  &$ \displaystyle {  x_1 -x_2=(x_3-x_2) \frac{    s_1^2+    s_3^2-s_2^2 }{ 2
   s_3^2} }$  & $ \displaystyle { \tanh(x_1 -x_2)= \tanh(x_3-x_2) \frac{ \cosh  s_1 \cosh  s_3-\cosh s_2}{ \cosh s_1
 \sinh s_3  \tanh s_3} }$  \\[2pt]

$ \displaystyle {   y_1  -y_2 = (y_3-y_2)  \frac{ \cos s_2- \cos s_1 \cos s_3 }{ 
\sin^2s_3} }$  &$ \displaystyle {   y_1 -y_2=(y_3-y_2) \frac{    s_1^2+    s_3^2-s_2^2 }{ 2
   s_3^2}   }$  & $ \displaystyle {  y_1  -y_2 = (y_3-y_2)  \frac{ \cosh s_1 \cosh s_3-\cosh s_2 }{ 
\sinh^2s_3}  }$  \\[8pt]
$ \displaystyle { \qquad \pm 2 \tan(x_3-x_2)    \frac{ \cos(\frac{s_1}{2})\cos(\frac{s_2}{2})\cos(\frac{s_3}{2})
 {A}  }{\sin s_3\tan s_3}  }$  &$ \displaystyle { \qquad\qquad\quad \pm 2  (x_3-x_2)    \frac{  
  {A}  }{  s_3^2}  }$  & $ \displaystyle { \qquad \pm 2 \tanh(x_3-x_2)    \frac{ \cosh(\frac{s_1}{2})\cosh(\frac{s_2}{2})\cosh(\frac{s_3}{2})
A }{\sinh s_3\tanh   s_3}  }$  \\[12pt]

$\bullet$ Anti-de Sitter space $\mathbf S^2_{[+],-}={\bf AdS}^{1+1}$ & $\bullet$ Minkowskian plane $\mathbf  S^2_{[0],-}={\bf M}^{1+1}$
&$\bullet$ De Sitter  space $\mathbf  S^2_{[-],-}={\bf dS}^{1+1}$ \\[4pt] 

$ C=v_{3}v_{0}+\tfrac{1}{2}\bigl(  v_{1}^{2}-v_{2}^{2}-  v_3^2\bigr) $&
$ C=v_{3}v_{0}+\tfrac{1}{2}\bigl( v_{1}^{2}-v_{2}^{2} \bigr) $&
$ C=v_{3}v_{0}+\tfrac{1}{2}\bigl( v_{1}^{2}-v_{2}^{2}+v_3^2\bigr) $\\[2pt]
   
$F^{(2)}=1-\cos(x_1-x_2) \cosh y_1 \cosh y_2+ 
 \sinh y_1 \sinh y_2       $& $F^{(2)}=\tfrac{1}{2}\left[
(x_{1}-x_{2})^{2}-(y_{1}-y_{2})^{2} \right] $&
$F^{(2)}=  \cosh(x_1-x_2) \cos  y_1 \cos  y_2+
\sin y_1 \sin y_2-1 $\\[2pt]
 $ \qquad\, =1- \cos s_1  $&  $\qquad\, =\tfrac 12 s_1^2$& $\qquad\, =\cosh s_1 -1 $\\ 

$ \displaystyle { \tan(x_1 -x_2)=\tan(x_3-x_2) \frac{ \cos s_2-\cos  s_1 \cos  s_3}{ \cos s_1
 \sin s_3  \tan s_3} }$  &$ \displaystyle {  x_1 -x_2=(x_3-x_2) \frac{    s_1^2+    s_3^2-s_2^2 }{ 2
   s_3^2} }$ & $ \displaystyle { \tanh(x_1 -x_2)= \tanh(x_3-x_2) \frac{ \cosh  s_1 \cosh  s_3-\cosh s_2}{ \cosh s_1
 \sinh s_3  \tanh s_3} }$  \\[8pt]
$ \displaystyle { \qquad \pm 4 \sinh(y_3-y_2)    \frac{ \cos(\frac{s_1}{2})\cos(\frac{s_2}{2})\cos(\frac{s_3}{2})
\sinh(\frac{A}{2}) }{\cos s_1\sin^2 s_3}  }$  &$ \displaystyle { \qquad\qquad\quad \pm 2  (y_3-y_2)    \frac{  
  {A}  }{  s_3^2}  }$  & $ \displaystyle { \qquad \pm 4 \sin(y_3-y_2)    \frac{ \cosh(\frac{s_1}{2})\cosh(\frac{s_2}{2})\cosh(\frac{s_3}{2})
\sinh(\frac{A}{2}) }{\cosh s_1\sinh^2 s_3}  }$  \\[8pt]

$ \displaystyle {  \sinh(y_1  -y_2)=\sinh(y_3-y_2)  \frac{ \cos s_2- \cos s_1 \cos s_3 }{ 
\sin^2 s_3} }$  &$ \displaystyle {   y_1 -y_2=(y_3-y_2) \frac{    s_1^2+    s_3^2-s_2^2 }{ 2
   s_3^2}   }$  & $ \displaystyle { \sin(y_1  -y_2)=\sin(y_3-y_2)  \frac{ \cosh s_1 \cosh s_3-\cosh s_2 }{ 
\sinh^2s_3}  }$  \\[8pt]

$ \displaystyle { \qquad \pm 4 \tan(x_3-x_2)    \frac{ \cos(\frac{s_1}{2})\cos(\frac{s_2}{2})\cos(\frac{s_3}{2})
\sinh(\frac{A}{2}) }{\sin s_3\tan s_3}  }$  &$ \displaystyle { \qquad\qquad\quad \pm 2  (x_3-x_2)    \frac{  
  {A}  }{  s_3^2}  }$  & $ \displaystyle { \qquad \pm 4 \tanh(x_3-x_2)    \frac{ \cosh(\frac{s_1}{2})\cosh(\frac{s_2}{2})\cosh(\frac{s_3}{2})
\sinh(\frac{A}{2}) }{\sinh s_3\tanh   s_3}  }$  \\[10pt]

\hline
\end{tabular}\hfill}
\end{table}

\end{landscape}

\noindent
The area of the triangle $\triangle  Q_1Q_2Q_{3}$ is given by
\be
 A=\frac{ \Sk_{\k_1}(s_1) \Sk_{\k_1}(s_3)}{2 \Ck_{\kk_1}(\frac{s_1}{2})\Ck_{\kk_1}(\frac{s_2}{2})\Ck_{\kk_1}(\frac{s_3}{2}) }\,\alpha .
\label{gr}
\ee
Then, from (\ref{gq}) we find that
$$
 y_1-y_2=(y_3-y_2)\frac{ \Sk_{\k_1}(s_1)}{ \Sk_{\k_1}(s_3)}+ \Sk_{\k_1}(s_1) \, \alpha .
$$
 By substituting $\Sk_{\k_1}(s_1)$ (\ref{gp}) in the first term and $\alpha$ from (\ref{gr}) in the second one, we get
 $$
  y_1-y_2=(y_3-y_2)\frac{\Ck_{\k_1}(s_2)-\Ck_{\k_1}(s_1)\Ck_{\k_1}(s_3)}{\k_1 \Sk^2_{\k_1}(s_3)}+ \frac{2 \Ck_{\kk_1}(\frac{s_1}{2})\Ck_{\kk_1}(\frac{s_2}{2})\Ck_{\kk_1}(\frac{s_3}{2}) }{ \Sk_{\k_1}(s_3)} \, A ,
 $$
 which, by introducing $\Tk_{\k_1}(x_3-x_2)$ (\ref{gp}), leads to the solution $y_1^+$ (\ref{gpp}).

Finally, we would like to recall that for the   Lorentzian spaces with $\k_2=-1/c^2$ (third row of the tables), ${\mathbf
S}^2_{[\k_1],-}$, the  triangle $\triangle  Q_1Q_2Q_{3}$ is a time-like one, that is, with elliptic distances $s_i$ for $\mathbf {AdS^{1+1}}$ $(\k_1>0)$  and hyperbolic  ones for $\mathbf {dS^{1+1}}$ $(\k_1<0)$. Any orthogonal geodesic to a time-like one  is   space-like, so the distances $y_1-y_2$ and $y_3-y_2$ are hyperbolic in $\mathbf {AdS^{1+1}}$ and elliptic in $\mathbf {dS^{1+1}}$. In fact, both spaces are related through the interchange of time- and space-like geodesics. By contrast, in the   Riemannian   spaces with $\k_2>0$ (first row of the tables), ${\mathbf
S}^2_{[\k_1],+}$, there is only one type of distance, elliptic for $\mathbf {S^{2}}$ and hyperbolic for $\mathbf {H^{2}}$.


\section{Concluding remarks}

We have achieved the first LH systems on 2D Riemannian, Lorentzian, and Newtonian spaces along with their constants of motion and superposition rules by following a  geometrical CK approach.  The graded contraction procedure based upon the two    parameters $(\k_1,\k_2)$   (curvature and signature) provides a clear description of the relationships among 
all the structures involved, which have been explicitly shown   in tables~\ref{table1} and \ref{table2} for each specific space.

In this framework, some natural open problems arise, which could be expected to be solved by applying   similar geometrical techniques, namely:

 \begin{itemize}
 	
 	\item All of these 2D LH systems could be extended to higher dimensions    by starting from the known  isometries on these spaces. The main point to be analyzed in this construction is the role that   higher-order Casimirs  play in relation with constants of motion. Recall that    quasi-othogonal algebras,  in any dimension, are always endowed with a 
 	second-order Casimir related to the Killing--Cartan form; in   the 2D  case this is the only one. Nevertheless, in   three dimensions there is another  third-order Casimir,   in the 4D case   there is a fourth-order one, etc. (see~\cite{casimir}).

 	\item The class P$_2\simeq \mathfrak{sl}(2)\simeq \mathfrak{so}(2,1)$ of the classification of 2D Euclidean LH systems is spanned by the following vector fields in Cartesian coordinates $(x,y)$~\cite{BBHLS, BHLS}:
 	$$
 	X_1:=\frac{\partial}{\partial x} ,\qquad X_2:=x\,\frac{\partial}{\partial x}+y\,\frac{\partial}{\partial y} ,\qquad X_3:=(x^2-y^2)\frac{\partial}{\partial x}+2xy\,\frac{\partial}{\partial y} .
 	$$
 	Remarkably, these vector fields are conformal symmetries of the Euclidean plane $\mathbb R^2$. In particular, $X_1$ is the translation along the axis $x$, the vector field $X_2$ is a dilatation, and $X_3$ is a specific conformal transformation related to the $x$-axis. They also close on a Lie subalgebra of the conformal 
 	Euclidean algebra $ \mathfrak{so}(3,1)$. Moreover, such vector fields are also Hamiltonian vector fields relative to a symplectic form \cite{BBHLS}.
 	This suggests us to make use of the known conformal symmetries on the CK spaces~\cite{HS02} to develop the `curved' counterparts of the LH systems of class  P$_2$. The physical relevance of this problem is due to the fact that P$_2$-LH systems underly the complex Riccati equation and some Milne--Pinney and Kummer--Schwarz equations~\cite{BHLS}.

	\item The class  I$_4\simeq \mathfrak{sl}(2)\simeq \mathfrak{so}(2,1)$ is spanned by the vector fields given by~\cite{BBHLS, BHLS}:
	$$
	X_1:=\frac{\partial}{\partial x} +\frac{\partial}{\partial y}  ,\qquad X_2:=x\,\frac{\partial}{\partial x}+y\,\frac{\partial}{\partial y} ,\qquad X_3:= x^2   \frac{\partial}{\partial x}+y^2 \frac{\partial}{\partial y} .
	$$
	Each $X_i$ has also a clear interpretation as conformal symmetries on the Euclidean line $\mathbb R$: $\partial_x+\partial_y$ is a translation, $x\partial_x+y\partial_y$ is a dilation and $x^2  {\partial_x}+y^2{\partial_y}$ is a conformal transformation \cite{GL17}. Therefore, the consideration of the conformal algebra on the 1D sphere $\mathbf S^1$ or hyperbolic line  $\mathbf H^1$ may lead to the `curved'  analog of the  I$_4$-LH systems. We recall that 
	I$_4$ covers the so-called  split-complex Riccati and coupled Riccati equations as well as some Milne--Pinney and Kummer--Schwarz equations (non-diffeomorphic to those of class P$_2$)~\cite{BHLS}.
	
	\item Finally, we also stress that the obtention of curved LH systems for the classes P$_2$ and  I$_4$ could further provide a curved oscillator system with a time-dependent  frequency and with a centrifugal or Winternitz term.  
	
\end{itemize}

These problems are currently under investigation.


\section*{Acknowledgments}

The research of  F.J.~Herranz was partially   supported by the Spanish Ministerio de Econom{\'{\i}}a y Competitividad     (MINECO) under grant MTM2013-43820-P, by   the grant MTM2016-79639-P (AEI/FEDER, UE) and   by   the Spanish Junta de Castilla y Le\'on  under grants BU278U14 and VA057U16.    J. de Lucas   acknowledges funding from the Polish National Science Centre  under   grant  MAESTRO (DEC-2012/06/A/ST1/00256).  M. Tobolski acknowledges partial support from the program Fizyka Plus with project number POKL.04.01.02-00-034/11-WF-37-18/13 carried out at the Faculty of Physics of the University of Warsaw and cofinanced by the European Union.


\end{document}